\documentclass[12pt,english]{article}

\usepackage[T1]{fontenc}
\usepackage[utf8]{inputenc}
\usepackage[dvipsnames]{xcolor}
\usepackage{babel}
\usepackage{amsthm}
\usepackage{amssymb}
\usepackage{mathtools,bbm,dsfont}
\usepackage{mathrsfs}
\usepackage{graphicx}
\usepackage{float}
\PassOptionsToPackage{normalem}{ulem}
\usepackage{etoolbox}
\usepackage{subcaption}
\usepackage{float}
\usepackage{soul}
\usepackage{cancel}
\usepackage{wasysym}
\usepackage{ulem}
\usepackage{comment}

\allowdisplaybreaks[1]

\usepackage{array,longtable}

\makeatletter

\usepackage{jheppub}
\def\@fpheader{\relax}
\allowdisplaybreaks[1]

\makeatother

\begin{document}

\title{A rotating GUP black hole: metric, shadow, and bounds on quantum parameters}


\author[a]{Federica Fragomeno,}
\author[b,e]{Samantha Hergott,}
\author[a,c,d]{Saeed Rastgoo,}
\author[a]{Evan Vienneau,}
\affiliation[a]{Department of Physics, University of Alberta, Edmonton, Alberta T6G 2G1, Canada} 
\affiliation[b]{Department of Physics and Astronomy, York University, Toronto, Ontario M3J 1P3, Canada}
\affiliation[c]{Department of Mathematical and Statistical Sciences, University of Alberta, Edmonton, Alberta T6G 2G1, Canada}
\affiliation[d]{Theoretical Physics Institute, University of Alberta, Edmonton, Alberta T6G 2G1, Canada}
\affiliation[e]{Perimeter Institute for Theoretical Physics, Waterloo, ON N2L 2Y5, Canada}
\emailAdd{ffragome@ualberta.ca}
\emailAdd{sherrgs@yorku.ca}
\emailAdd{srastgoo@ualberta.ca}
\emailAdd{eviennea@ualberta.ca}










\abstract{Recently, for the first time, a metric of a static spherically symmetric generalized uncertainty inspired quantum black hole was derived. We apply the modified Newman-Janis algorithm to this metric and derive its rotating counterpart. We show that this metric has all the correct limits, while due to Newman-Janis side effects, the singularity which was resolved in the static case, is introduced back into the model. However, the slowly-rotating limit of this black hole is singularity-free. Furthermore, we show that the presence of quantum parameters modifies the location of the horizons, temperature, and entropy of the black hole, and allows the existence of naked singularities even if the ratio of the spin parameter to mass of the black hole is less than unity. Finally, by computing the shadow parameters of this black hole and comparing them with data from the Event Horizon Telescope for both M87* and Sgr A*, we set bounds on one of the quantum parameters of the model, and show that there is a limit on the angular momentum of M87* if this model is valid.}







\maketitle

\section{Introduction}

Black holes provide one of the most promising arenas to test quantum-gravity-inspired modifications of classical general relativity. Several approaches to quantum gravity have put forward models of quantum black holes \cite{bambi2023regular} suggesting corrections that can soften or resolve the singularity in static, collapsing, and rotating geometries. There are also model-independent proposals showing these features for fully dynamical cases involving matter \cite{Hergott:2022hjm, Hergott:2025elg}. These effects are also present in lower dimensional black holes (see, e.g., \cite{Gambini:2009vp, Corichi:2016nkp}). In this context, generalized uncertainty principle (GUP) deformations offer a phenomenological framework in which short-distance quantum effects can be encoded in modified black hole metrics and observables~\cite{Ali:2010yn, Blanchette:2021vid, Anacleto:2021qoe, Bosso:2023aht, Bosso:2023fnb, Rastgoo:2022mks, Bosso:2020ztk}. 

The phenomenological framework of GUP is strongly motivated by several independent approaches to quantum gravity, which suggest the existence of a minimal length at the Planck scale. In string theory, the study of ultra-high-energy string scattering reveals that strings expand at energy scales approaching the Planck mass, preventing sub-Planckian spatial resolution and naturally modifying the Heisenberg uncertainty relation to $\Delta x \sim \hbar/\Delta p + \alpha' \Delta p$ \cite{Amati:1988tn, Gross:1987ar}. Similarly, in the context of non-commutative geometry, spacetime coordinates are promoted to non-commuting operators, $[x^\mu, x^\nu] = i \theta^{\mu\nu}$, which intrinsically bounds the simultaneous measurement of spacetime events and implies a minimal length scale \cite{Snyder:1946qz, Connes:1994yd}. Related minimum-length features also emerge in loop quantum gravity through the discrete spectra of geometric operators like area and volume \cite{Rovelli:1994ge}. Because these diverse fundamental theories converge on the necessity of modifying the standard uncertainty principle at high energies, the GUP serves as a powerful, model-independent tool. It allows us to effectively capture these universal short-distance quantum gravity effects and incorporate them into the metric and thermodynamics of black holes without requiring the full mathematical machinery of the underlying fundamental theories \cite{Maggiore:1993rv, Kempf:1994su}.

Recently, a static spherically symmetric improved GUP black hole metric, possessing two quantum parameters $Q_b$ and $Q_c$, was derived systematically \cite{Fragomeno:2024tlh, Gingrich:2024mgk}. In that construction, $Q_b$ primarily controls near-horizon and exterior quantum corrections, whereas $Q_c$ mainly affects the deep interior structure. A natural next step is to obtain the rotating counterpart of this static solution, since astrophysical black holes are expected to possess nonzero angular momentum. Hence, introducing rotation is crucial for confronting the model with observations.

To construct the rotating geometry, one method is to employ the Newman-Janis (NJ) algorithm \cite{Newman:1965tw}, or modified variants of it \cite{Azreg-Ainou:2014nra,Azreg-Ainou:2014aqa,Azreg-Ainou:2014pra}. However, the NJ procedure is known to be subtle beyond the original Kerr construction when the original equations of motion of the underlying theory are not known: depending on the seed metric and implementation details, it may generate geometries with nontrivial pathologies, and regularity properties of the static seed are not guaranteed to persist~\cite{HansenYunes2013,KamenshchikPetriakova2023,ShaoChenChen2021,NevesSaa2014}. This motivates a careful analysis of the asymptotic and near-singularity limits, curvature behavior, and horizon structure in any NJ-generated rotating extension.

In this work, we apply a modified NJ algorithm to the improved static GUP black-hole metric  introduced in \cite{Fragomeno:2024tlh} and derive its rotating version. We show that the resulting spacetime has the appropriate classical and nonrotating limits. We then study the horizon structure and extremality conditions, and show that the quantum effects can shift horizon radii and allow naked singularity configurations even for $\frac{a}{M}<1$. We also compute thermodynamic quantities and find that quantum effects lower both the temperature and entropy compared with the classical rotating case.

We also investigate whether the model has a singularity, find that it does, and identify similar pathologies to those reported previously \cite{HansenYunes2013,KamenshchikPetriakova2023,ShaoChenChen2021,NevesSaa2014}. However, we show that the slowly-rotating limit is singularity-free.

Finally, we investigate observational signatures by computing the shadow of the rotating GUP black hole and comparing it with Event Horizon Telescope (EHT) measurements for M87* and Sgr A* \cite{EventHorizonTelescope:2019dse,EventHorizonTelescope:2019pgp,EventHorizonTelescope:2022wkp}. From this comparison, we place bounds on the quantum parameter $Q_b$ and identify an upper bound on the allowed spin of M87* within this model. These results provide a direct bridge between quantum gravity-motivated black hole geometry and current horizon scale observations.

The paper is organized as follows. In Sec. \ref{Sec:NJ-review} we review the modified NJ algorithm used in our construction and present the general rotating metric associated to a static black hole. In Sec. \ref{Sec:Rot-metric} we present the rotating metric of the GUP rotating black hole and analyze its consistency by studying the asymptotic and classical limits. We also study the horizons, extremality condition, and thermodynamic properties of the black hole. Furthermore, we compute the Kretschmann scalar in both the fully rotating and slowly-rotating cases, and also find the affine distance to the $r=0$ region. Sec. \ref{Sec:shadow} is the phenomenological heart of the paper in which we compute the shadow of the black hole and derive constraints on its quantum parameters by comparing the model to the EHT data from Sgr A* and M87*. Finally, we summarize our results and conclude in Sec. \ref{sec:Conclusion}.

\section{Brief Overview Of The Modified Newman-Janis Algorithm\label{Sec:NJ-review}}

The NJ algorithm \cite{Newman:1965tw} was introduced
as a means to derive the Kerr metric from the static spherically symmetric
Schwarzschild solution. Recently, a modified version of it was presented~\cite{Azreg-Ainou:2014nra,Azreg-Ainou:2014aqa,Azreg-Ainou:2014pra} in which the explicit complexification
of the coordinates is avoided and instead certain other conditions are introduced to
arrive at the final result (see below). Here we briefly
describe this modified algorithm.

Consider the generic static spherically symmetric metric 
\begin{equation}
ds^{2}=g_{00}dt^{2}+g_{11}dr^{2}+g_{22}d\Omega^{2},
\end{equation}
where $d\Omega^{2}=d\theta^{2}+\sin^{2}(\theta)d\phi^{2}$ and hence
$g_{33}=g_{22}\sin^{2}(\theta)$. In order to derive a rotating counterpart
of this metric, one first transforms to the outgoing Eddington-Finkelstein
(EF) coordinates
\begin{equation}
u=t-r^{*}=t-\int\sqrt{-\frac{g_{11}}{g_{00}}}dr,
\end{equation}
where $r^{*}$ is the radial tortoise coordinate. The metric is now
\begin{align}
ds^{2}= & g^{(\text{EF})}_{\tilde{0}\tilde{0}}du^{2}+2g^{(\text{EF})}_{\tilde{0}\tilde{1}}dudr+g^{(\text{EF})}_{\tilde{2}\tilde{2}}d\Omega^{2}\nonumber \\
= & g_{00}du^{2}+2\text{sgn}\left(g_{00}\right)\sqrt{-g_{00}g_{11}}dudr+g_{22}d\Omega^{2},
\end{align}
where $g_{\tilde{\mu}\tilde{\nu}}$ are the components of the metric
in the EF coordinates. Using the Newman-Penrose formalism \cite{Newman:1961qr},
the inverse of this metric
\begin{align}
g^{(\text{EF})\tilde{0}\tilde{1}}= & \frac{1}{g^{(\text{EF})}_{\tilde{0}\tilde{1}}}, & g^{(\text{EF})\tilde{1}\tilde{1}}= & -\frac{g^{(\text{EF})}_{\tilde{0}\tilde{0}}}{\left[g^{(\text{EF})}_{\tilde{0}\tilde{1}}\right]^{2}}\\
g^{(\text{EF})\tilde{2}\tilde{2}}= & \frac{1}{g^{(\text{EF})}_{\tilde{2}\tilde{2}}}, & g^{(\text{EF})\tilde{3}\tilde{3}}= & \frac{1}{g^{(\text{EF})}_{\tilde{3}\tilde{3}}}
\end{align}
is now written in terms of four null tetrads $l^{\tilde{\mu}},\,n^{\tilde{\mu}},\,m^{\tilde{\mu}},\,\bar{m}^{\tilde{\mu}}$,
as
\begin{equation}
g^{(\text{EF})\tilde{\mu}\tilde{\nu}}=-l^{\tilde{\mu}}n^{\tilde{\nu}}-l^{\tilde{\nu}}n^{\tilde{\mu}}+m^{\tilde{\mu}}\bar{m}^{\tilde{\nu}}+m^{\tilde{\nu}}\bar{m}^{\tilde{\mu}}\label{eq:g-expand}
\end{equation}
where $l^{\tilde{\mu}},\,n^{\tilde{\mu}}$ are real and $m^{\tilde{\mu}}$
is complex with its complex conjugate being $\bar{m}^{\tilde{\mu}}$,
and they obey the normalization conditions
\begin{align}
g^{(\text{EF})}_{\tilde{\mu}\tilde{\nu}}l^{\tilde{\mu}}n^{\tilde{\nu}}= & -1, & g^{(\text{EF})}_{\tilde{\mu}\tilde{\nu}}m^{\tilde{\mu}}\bar{m}^{\tilde{\nu}}= & 1
\end{align}
with the rest of the inner products vanishing. From Eq. \eqref{eq:g-expand}
and the above normalization conditions (and the null property of the
tetrads) one can find them as
\begin{align}
l^{\tilde{\mu}}= & \delta^{\tilde{\mu}}_{\tilde{1}},\\
n^{\tilde{\mu}}= & \frac{1}{\sqrt{-g_{00}g_{11}}}\delta^{\tilde{\mu}}_{\tilde{0}}-\frac{1}{2g_{11}}\delta^{\tilde{\mu}}_{\tilde{1}},\\
m^{\tilde{\mu}}= & \frac{1}{\left(\sqrt{2g_{22}}\right)^{*}}\delta^{\tilde{\mu}}_{\tilde{2}}\pm\frac{i}{\sin\left(\theta\right)}\frac{1}{\left(\sqrt{2g_{22}}\right)^{*}}\delta^{\tilde{\mu}}_{\tilde{3}},\\
\bar{m}^{\tilde{\mu}}= & \frac{1}{\left(\sqrt{2g_{22}}\right)^{*}}\delta^{\tilde{\mu}}_{\tilde{2}}\mp\frac{i}{\sin\left(\theta\right)}\frac{1}{\left(\sqrt{2g_{22}}\right)^{*}}\delta^{\tilde{\mu}}_{\tilde{3}}.
\end{align}
In the original NJ algorithm for the Schwarzschild black hole, $\frac{1}{\left(\sqrt{2g_{22}}\right)^{*}}=\frac{1}{\sqrt{2}r^{*}}$,
and hence one complexifies $r$ while keeping $n^{\tilde{\mu}}$
real. In the modified approach one avoids that step (see below for the conditions that replace this). Next, one introduces the angular momentum into
the metric by performing a deformation transformation
\begin{align}
r^{\prime}= & r+ia\cos\left(\theta\right), & \theta^{\prime}= & \theta,\\
u^{\prime}= & u-ia\cos\left(\theta\right), & \phi^{\prime}= & \phi.
\end{align}
At the same time, one assumes that under this transformation, the $g_{\mu\nu}(r)$
components become $\breve{g}_{\mu\nu}(r,\theta,a)$. The condition
replacing complexification is $\lim_{a\to0}\breve{g}_{\mu\nu}=g_{\mu\nu}$. 
The above transformation results in $g^{\mu^{\prime}\nu^{\prime}}$
from which one can find $g_{\mu^{\prime}\nu^{\prime}}$ as
\begin{align}
ds^{\prime(\text{EF})}= & \breve{g}_{00}du^{\prime2}-2\sqrt{-\breve{g}_{00}\breve{g}_{11}}du^{\prime}dr^{\prime}-2a\sin^{2}\left(\theta\right)\left(\breve{g}_{00}+\sqrt{-\breve{g}_{00}\breve{g}_{11}}\right)du^{\prime}d\phi\nonumber \\
 & +2a\sin^{2}\left(\theta\right)\sqrt{-\breve{g}_{00}\breve{g}_{11}}dr^{\prime}d\phi\nonumber \\
 & +\breve{g}_{22}d\theta^{2}+\sin^{2}\left(\theta\right)\left[\breve{g}_{22}+a^{2}\sin^{2}\left(\theta\right)\left(\breve{g}_{00}+2\sqrt{-\breve{g}_{00}\breve{g}_{11}}\right)\right]d\phi^{2}.
\end{align}
Note that all instances of $g_{\mu\nu}(r)$ are replaced by $\breve{g}_{\mu\nu}(r,\theta,a)$. One now makes
a transformation
\begin{align}
du^{\prime}= & d\bar{t}+\lambda\left(\bar{r}\right)d\bar{r}, & dr^{\prime}= & d\bar{r}\\
d\phi= & d\bar{\phi}+\chi\left(\bar{r}\right)d\bar{r}, & d\theta= & d\bar{\theta}
\end{align}
to the Boyer-Lindquist (BL) coordinates. The reason $\lambda$ and $\chi$ depend only on $r$ is that if they also depended on $\theta$ and $\phi$, the transformation would be invalid, and the NJ algorithm would fail to transform the metric into BL form. The unknowns $\lambda\left(\bar{r}\right)$ and $\chi\left(\bar{r}\right)$
are found using three criteria: 1) The BL conditions $g_{\bar{t}\bar{r}}=0=g_{\bar{r}\bar{\phi}}$.
2) the dependence of $\lambda\left(\bar{r}\right)$ and $\chi\left(\bar{r}\right)$ is solely on $\bar{r}$. 3) $\lim_{a\to0}\breve{g}_{\mu\nu}=g_{\mu\nu}$.
Finding these unknowns and replacing them back into the metric after
the above transformation yields the rotating metric components
\begin{align}
g_{\bar{0}\bar{0}}= & -\frac{\breve{g}_{22}\left[\frac{g_{22}}{g_{11}}+a^{2}\cos^{2}\left(\theta\right)\right]}{\left[\frac{g_{22}}{\sqrt{-g_{00}g_{11}}}+a^{2}\cos^{2}\left(\theta\right)\right]^{2}},\\
g_{\bar{0}\bar{3}}= & -a\sin^{2}\left(\theta\right)\breve{g}_{22}\frac{\frac{g_{22}}{\sqrt{-g_{00}g_{11}}}-\frac{g_{22}}{g_{11}}}{\left[\frac{g_{22}}{\sqrt{-g_{00}g_{11}}}+a^{2}\cos^{2}\left(\theta\right)\right]^{2}},\\
g_{\bar{1}\bar{1}}= & \frac{\breve{g}_{22}}{\frac{g_{22}}{g_{11}}+a^{2}},\\
g_{\bar{2}\bar{2}}= & \breve{g}_{22},\\
g_{\bar{3}\bar{3}}= & \breve{g}_{22}\sin^{2}\left(\theta\right)\left[1+a^{2}\sin^{2}\left(\theta\right)\left(\frac{a^{2}\cos^{2}\left(\theta\right)-\frac{g_{22}}{g_{11}}+2\frac{g_{22}}{\sqrt{-g_{00}g_{11}}}}{\left[\frac{g_{22}}{\sqrt{-g_{00}g_{11}}}+a^{2}\cos^{2}\left(\theta\right)\right]^{2}}\right)\right].
\end{align}
Notice that this algorithm does not fix $\breve{g}_{22}$. This can
be fixed by requiring that the classical and nonrotating limits
match their expected limits (see below), which yields
\begin{equation}
\breve{g}_{22}=g_{22}+a^{2}\cos^{2}\left(\theta\right).
\end{equation}
It is worth noting that there exists another method of obtaining a
slowly-rotating metric from the static one \cite{Kumar:2019uwi, Balali:2024mtt},
which in our case can simply be obtained by taking the $a^{2}\to0$ limit of the above
metric but keeping the terms first order in $a$.

\section{A GUP Rotating Black Hole Metric\label{Sec:Rot-metric}}

\subsection{The Full Metric}
To incorporate quantum gravitational effects at the semi-classical level using GUP, the standard phase space structure is deformed by modifying the fundamental Poisson brackets. Specifically, one can introduce quadratic corrections in the configuration variables (in this case Ashtekar connection components) $b,\, c$, such that, $\{b,p_b\}=1+\beta_b b^2$ and $\{c,p_c\}=1+\beta_c c^2$. Here $p_b,\, p_c$ are the momenta conjugate to $b,\, c$, and $\beta_b,\, \beta_c$ are the constant deformation parameters of GUP associated to $b$ and $c$ sectors. These modifications of the algebra, however, lead to incorrect asymptotic limit. To resolve this issue, in \cite{Fragomeno:2024tlh} the authors applied the ``improved'' method (borrowed from LQG) to the deformations, in which the constant deformation parameters $\beta_b,\, \beta_c$ are made momentum dependent: $\beta_b\to \bar{\beta}_b=\beta_b/p^{2}_{b}$ and $\beta_c\to \bar{\beta}_c=\beta_c/p^{2}_{c}$. Physically this can be interpreted as deformations that are not just quantum, but are quantum gravitational, since $p_b,\, p_c$ are the metric components.

Following this procedure a metric of the improved GUP black hole can be derived systematically as ~\cite{Fragomeno:2024tlh}
\begin{align}
g_{00}(r)= & -\left(1+\frac{Q_{b}}{r^{2}}\right)\left(1+\frac{Q_{c}R^{2}_{s}}{4r^{8}}\right)^{-\frac{1}{4}}\left(1-\frac{R_{s}}{\sqrt{r^{2}+Q_{b}}}\right)\label{eq:GUP-non-rot-00},\\
g_{11}(r)= & \left(1+\frac{Q_{c}R^{2}_{s}}{4r^{8}}\right)^{\frac{1}{4}}\left(1-\frac{R_{s}}{\sqrt{r^{2}+Q_{b}}}\right)^{-1},\label{eq:GUP-non-rot-11}\\
g_{22}(r)= & r^{2}\left(1+\frac{Q_{c}R^{2}_{s}}{4r^{8}}\right)^{\frac{1}{4}},\label{eq:GUP-non-rot-22}\\
g_{33}(r)= & r^{2}\left(1+\frac{Q_{c}R^{2}_{s}}{4r^{8}}\right)^{\frac{1}{4}}\sin^{2}\left(\theta\right),\label{eq:GUP-non-rot-33}
\end{align}
Here, $Q_b\sim\beta_b$ and $Q_c\sim\beta_c$ are the two quantum parameters of the model. The former introduces quantum gravity effects in the exterior and near the horizon, while the latter is responsible for such effects in the interior and particularly singularity resolution.

This metric has correct classical ($Q_b\to 0,\, Q_b\to 0$) and asymptotic ($r\to\infty$) limits and also resolves the classical singularity.

\begin{figure}[tp]
\centering \includegraphics[width=0.8\linewidth]
{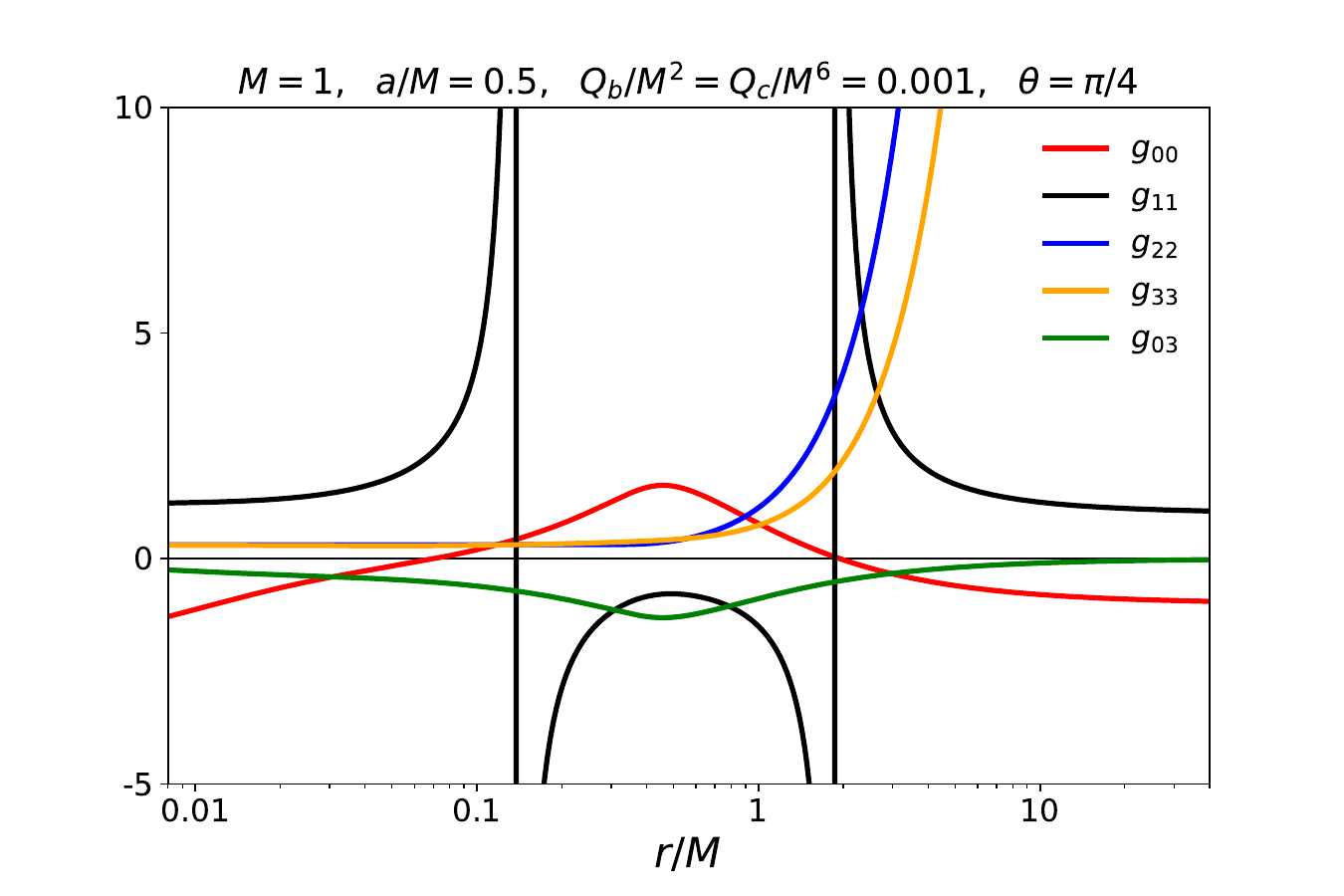}
\caption{Plot of rotating GUP metric components in Schwarzschild coordinates with the given values of parameters on the top of the plot.}\label{metric_comps}
\end{figure}

Using the above metric and following the method presented in the previous section, the NJ-rotated version of the static GUP metric is obtained as
\begin{align}
g_{\bar{0}\bar{0}}= & -\frac{\rho^{2}}{\Sigma^{2}}\left(\Sigma-\varLambda\right),\label{eq:GUP-rot-00}\\
g_{\bar{0}\bar{3}}= & -\frac{a\sin^{2}\left(\theta\right)\rho^{2}}{\Sigma^{2}}\varLambda,\label{eq:GUP-rot-03}\\
g_{\bar{1}\bar{1}}= & \frac{\rho^{2}}{\Delta},\label{eq:GUP-rot-11}\\
g_{\bar{2}\bar{2}}= & \rho^{2},\label{eq:GUP-rot-22}\\
g_{\bar{3}\bar{3}}= & \frac{\rho^{2}\sin^{2}\left(\theta\right)}{\Sigma^{2}}\left[\Sigma^{2}+a^{2}\sin^{2}\left(\theta\right)\left(\Sigma+\varLambda\right)\right],\label{eq:GUP-rot-33}
\end{align}
where
\begin{align}
\rho^{2}= & \breve{g}_{22}=g_{22}+a^{2}\cos^{2}\left(\theta\right)=r^{2}\left(1+\frac{Q_{c}R^{2}_{s}}{4r^{8}}\right)^{\frac{1}{4}}+a^{2}\cos^{2}\left(\theta\right),\label{eq:rho-def}\\
\Delta= & \frac{g_{22}}{g_{11}}+a^{2}=r^{2}\left(1-\frac{R_{s}}{\sqrt{r^{2}+Q_{b}}}\right)+a^{2},\label{eq:Delta-def}\\
\Sigma= & \frac{g_{22}}{\sqrt{-g_{00}g_{11}}}+a^{2}\cos^{2}\left(\theta\right)=\left(\rho^{2}-a^{2}\cos^{2}\left(\theta\right)\right)\left(1+\frac{Q_{b}}{r^{2}}\right)^{-\frac{1}{2}}+a^{2}\cos^{2}\left(\theta\right),\label{eq:Sigma-def}\\
\varLambda= & \Sigma-\Delta+a^{2}\sin^{2}\left(\theta\right).\label{eq:Lambda-def}
\end{align}
A plot of these metric components for certain values of parameters is
depicted in Fig.~\ref{metric_comps}.

We notice from \eqref{eq:rho-def}-\eqref{eq:Sigma-class} that for
the classical limit where $Q_{b},\,Q_{c}\to0$ we get
\begin{align}
\rho^{2}_{\text{class}}= & r^{2}+a^{2}\cos^{2}\left(\theta\right),\label{eq:rho-class}\\
\Delta_{\text{class}}= & r^{2}\left(1-\frac{R_{s}}{r}\right)+a^{2},\label{eq:Delta-class}\\
\Sigma_{\text{class}}= & \rho^{2}.\label{eq:Sigma-class}
\end{align}
In the same way, for the nonrotating (and also slowly-rotating) limit
we obtain
\begin{align}
\rho^{2}_{\text{NR}}= & r^{2}\left(1+\frac{Q_{c}R^{2}_{s}}{4r^{8}}\right)^{\frac{1}{4}},\label{eq:rho-NR}\\
\Delta_{\text{NR}}= & r^{2}\left(1-\frac{R_{s}}{\sqrt{r^{2}+Q_{b}}}\right),\label{eq:Delta-NR}\\
\Sigma_{\text{NR}}= & \rho^{2}\left(1+\frac{Q_{b}}{r^{2}}\right)^{-\frac{1}{2}},\label{eq:Sigma-NR}
\end{align}
and in the combined classical and nonrotating limit we get
\begin{align}
\rho^{2}_{\text{class-NR}}= & r^{2},\label{eq:rho-class-NR}\\
\Delta_{\text{class-NR}}= & r^{2}\left(1-\frac{R_{s}}{r}\right),\label{eq:Delta-class-NR}\\
\Sigma_{\text{class-NR}}= & r^{2}.\label{eq:Sigma-class-NR}
\end{align}
The above results show that the classical limit of the metric is the
Kerr metric, the nonrotating limit is the static GUP metric \eqref{eq:GUP-non-rot-00}-\eqref{eq:GUP-non-rot-33}
and the classical nonrotating case reduces to the Schwarzschild metric
as it should.

One can also check that the asymptotic expansion of the metric components
behave as they should. In particular, they reduce to the Kerr components for
$Q_{b},Q_{c}\to0$ and to the Schwarzschild case for $a\to0$ and $Q_{b},Q_{c}\to0$. 

\subsection{Horizons}

\begin{figure}[tp]
\centering \includegraphics[width=0.65\linewidth]{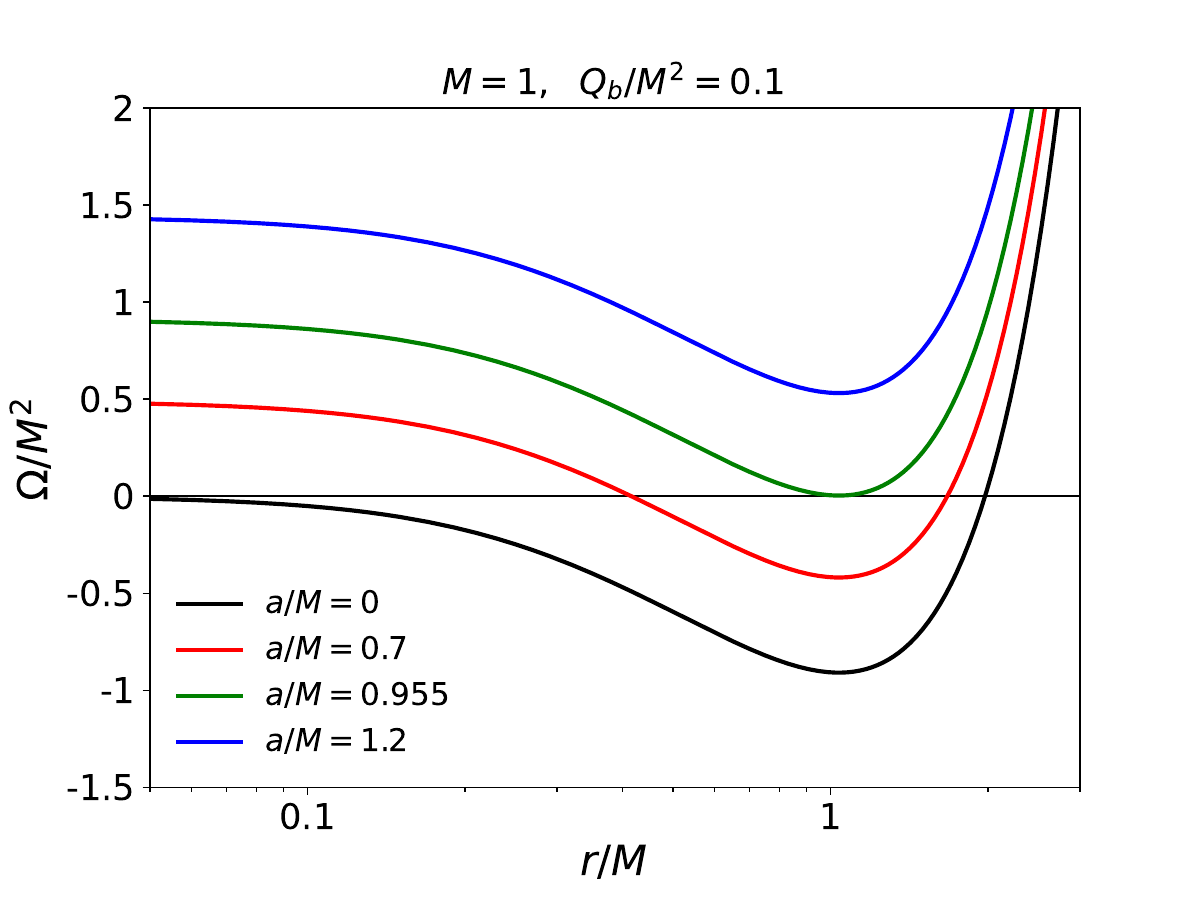}
\caption{Plot of the horizon function, $\Omega(r)$, for several values of the spin parameter $a$, with the given value of parameters on the top of the plot.}
\label{horizon}
\end{figure}

As usual, the horizons of such a black hole can be derived by setting
$g^{11}=0$, which corresponds to
\begin{equation}
\Delta\left(r\right)=r^{2}\left(1-\frac{R_{s}}{\sqrt{r^{2}+Q_{b}}}\right)+a^{2}=0.
\end{equation}
Since this is a quadratic equation and also as can be seen from Fig.
\ref{horizon}, $\Delta(r)$ admits at most two roots corresponding
to an outer and an inner horizon. The inner and outer horizon radii
can be obtained perturbatively by solving $\Delta=0$ to first order
in the parameter $Q_{b}$, 
\begin{align}
r_{-}= & \frac{1}{2}\left(R_{s}-\sqrt{R^{2}_{s}-4a^{2}}\right)+\frac{Q_{b}R_{s}}{R_{s}\left(\sqrt{R^{2}_{s}-4a^{2}}-R_{s}\right)+4a^{2}},\\
r_{+}= & \frac{1}{2}\left(R_{s}+\sqrt{R^{2}_{s}-4a^{2}}\right)-\frac{Q_{b}R_{s}}{R_{s}\left(\sqrt{R^{2}_{s}-4a^{2}}+R_{s}\right)-4a^{2}}.
\end{align}
As expected, in the limit $Q_{b}\rightarrow0$ these expressions smoothly
reduce to the Kerr horizons. The quantum parameter shrinks the outer
horizon and expands the inner horizon relative to their classical
values. Since both ends of the function $\Delta(r)$ are positive, i.e., $\Delta(r\to 0)\rightarrow a^{2}>0$ and
$\Delta(r\to\infty)\sim r^{2}+a^{2}>0$, the
horizons will only exist if there is a minimum of $\Delta(r)$ where $\Delta(r)\leq 0$. The extremal case happens when this root
just touches the $r$ axis, namely there is a degenerate root of $\Delta(r)$,
in which case the two horizons coincide. This condition means the
extremal horizon located at $r_{E}$ happens when 
\begin{align}
\Delta(r_{E}), & =0 & \Delta^{\prime}(r_{E}) & =0,
\end{align}
where
\begin{equation}
\Delta^{\prime}\left(r\right)=2r-\frac{R_{s}r\left(r^{2}+2Q_{b}\right)}{\left(r^{2}+Q_{b}\right)^{3/2}}.\label{eq:Delta-prime}
\end{equation}
The resulting extremality relations are 
\begin{align}
R_{s}= & \frac{2(r^{2}_{E}+Q_{b})^{\frac{3}{2}}}{r^{2}_{E}+2Q_{b}} & a^{2}_{E}= & \frac{r^{4}_{E}}{r^{2}_{E}+2Q_{b}}\approx\frac{R^{2}_{s}}{4}-2Q_{b},
\end{align}
where $a^{2}_{E}$ is the extremal spin parameter such that for $a^{2}<a^{2}_{E}$
there are two horizons, for $a^{2}=a^{2}_{E}$ there is one horizon
(extremal case), and for $a^{2}>a^{2}_{E}$ there are no horizons
(naked singularity). The above approximation is valid for $Q_b \ll R_s$. In this limit, we obtain
\begin{equation}
r_{E}\approx\sqrt{\frac{R^{2}_{s}}{4}-Q_{b}}.
\end{equation}
As a consistency check, in the classical limit, $Q_{b}\rightarrow0$,
we recover $a_{E}=\frac{R_{s}}{2}$, exactly as in Kerr. This suggests
that $Q_{b}$ lowers the maximum allowed spin for which horizons exist, as reflected in the condition $a_E^2 < r_E^2$.
As mentioned above, for spins exceeding this extremal value $a^{2}>a^{2}_{E}$,
the minimum of $\Delta(r_{E})$ remains positive and we do not have
any horizons.

It is worth noting that the horizon dynamics we observe here---specifically, the quantum parameter $Q_b$ shrinking the outer horizon and expanding the inner horizon---is a recurring geometric feature in various phenomenological models of quantum-corrected black holes. A prominent example is found in the asymptotic safety scenario of quantum gravity. In this framework, renormalization-group (RG) improved Schwarzschild geometries \cite{Bonanno:2000ep} and their rotating Kerr counterparts \cite{Reuter:2010xbd} feature a scale-dependent running Newton's constant. This quantum correction naturally reduces the size of the outer event horizon and generates an inner Cauchy horizon, which eventually merge to form an extremal configuration at a critical mass scale. Similar qualitative behavior appears in noncommutative geometry-inspired black holes, where the smearing of the central point mass due to a minimal length scale predictably shifts both horizons toward each other \cite{Nicolini:2005vd}. Furthermore, effective black hole models originating from loop quantum gravity, which resolve the classical singularity via quantum geometric constraints, also predict modifications to the horizon radii that parallel these effects \cite{Ashtekar:2018cay}. Therefore, the shift in horizon radii induced by $Q_b$ in our metric is highly consistent with the broader literature on quantum-gravitational deformations of classical spacetimes.

\begin{figure}[tp]
\centering \includegraphics[width=0.8\linewidth]{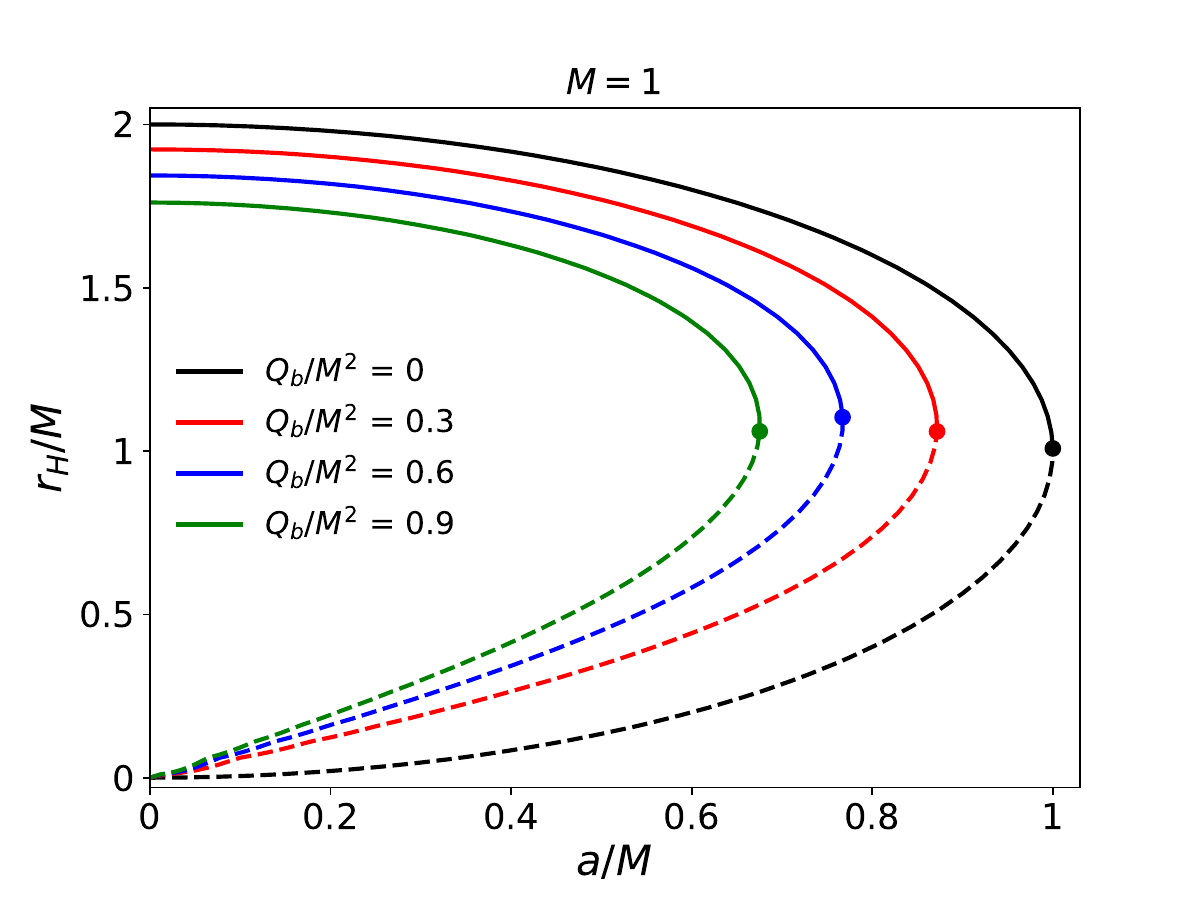}
\caption{Plot which shows the radial positions of inner (dashed colored lines) and outer (solid colored lines) horizons for an $M = 1$ black hole, as a function of the spin parameter $a$ for different values of the quantum parameter $Q_b$. Coloured points indicate the radii at which the inner and outer horizons merge to form one horizon, yielding an extremal black hole.}
\label{horizons}
\end{figure}

Figure~\ref{horizons} shows the positions of inner and outer horizons as a function of the spin parameter $a$ for various values of the quantum parameter $Q_b$, for a black hole with $M = 1$. As usual, for a nonrotating black hole, there is no inner horizon and the position of the single horizon decreases with increasing $Q_b$. For a rotating black hole, the distance between inner and outer horizons decrease with increasing $a$. Furthermore, the value of $a$ at which the inner and outer horizons merge into one decreases with increasing $Q_b$. Notably, a nonzero value of $Q_b$ allows for an extremal black hole to exist for $\frac{a}{M} < 1$, suggesting that in this model, naked singularities (if a singularity exists) can reasonably exist for specific parameter values. 

\subsection{Thermodynamics}

For stationary, axisymmetric spacetimes written in BL-like
coordinates, the surface gravity $\kappa$ at the outer event horizon
$r_{+}$ is determined by the radial derivative of the horizon function
$\Delta(r)$ as
\[
\kappa=\frac{\Delta^{\prime}\left(r_{+}\right)}{2\left(r^{2}_{+}+a^{2}\right)}.
\]\,
The Hawking temperature is directly proportional to this surface gravity
as
\begin{equation}
T_{H}=\frac{\kappa}{2\pi}=\frac{\Delta^{\prime}\left(r_{+}\right)}{4\pi\left(r^{2}_{+}+a^{2}\right)}.\label{eq:TH-general}
\end{equation}
Replacing \eqref{eq:Delta-prime} in the above expression, and noting that at the horizon 
\begin{equation}
\Delta\left(r_{+}\right)=0\Rightarrow\frac{R_{s}}{\sqrt{r^{2}_{+}+Q_{b}}}=\frac{r^{2}_{+}+a^{2}}{r^{2}_{+}},
\end{equation}
the temperature equation simplifies to 
\begin{equation}
T_{H}=\frac{2r_{+}}{4\pi\left(r^{2}_{+}+a^{2}\right)}-\frac{r^{2}_{+}+2Q_{b}}{4\pi r_{+}\left(r^{2}_{+}+Q_{b}\right)}.
\end{equation}
This reduces to the correct classical limit for $Q_{b}\to0$, and furthermore shows that the temperature is lower than in the classical case due to quantum effects. From \eqref{eq:TH-general}, it is clear that in the extremal case where $\Delta^{\prime}\left(r_{E}\right)=0=\Delta\left(r_{E}\right)$
at the horizon, the temperature is zero as it should be for an extremal
black hole.

For the entropy, we assume that the system still obeys 
\begin{equation}
S=\frac{A}{4},\label{eq:entropy-class}
\end{equation}
where $A$ is the area of the outer horizon. The area of $r_{+}$
is derived from the determinant of the $\theta-\phi$ block as
\begin{equation}
A=\int^{2\pi}_{0}\int^{\pi}_{0}\sqrt{g_{\theta\theta}g_{\phi\phi}-g^{2}_{\theta\phi}}d\theta d\phi=4\pi\left(r^{2}_{+}+a^{2}\right).
\end{equation}
Hence, the entropy is
\begin{equation}
S=\pi\left(r^{2}_{+}+a^{2}\right)\label{eq:entopy-quantum}
\end{equation}
which is smaller than the classical case given that $r_{+}$ is smaller
in the GUP case compared to its classical counterpart.

It is crucial to clarify the physical assumptions underlying the choice of entropy in Eq. \eqref{eq:entropy-class} leading to \eqref{eq:entopy-quantum}. Because the quantum-corrected metric considered here is phenomenological and not derived from a specific variational principle of a known effective field theory, we lack the explicit underlying action required to rigorously compute the exact Wald entropy \cite{Wald:1993nt}. Consequently, the strict area law $S = A/4$ utilized here must be understood as a leading-order approximation. Retaining the classical entropy while introducing quantum corrections to the mass, temperature, and spin implies that the exact thermodynamic first law, $dM = T dS + \Omega dJ$, may not be identically satisfied at all orders. In a fully consistent quantum thermodynamic framework, demanding the validity of the first law alongside a modified temperature naturally sources higher-order quantum corrections to the entropy itself. Such corrections, which frequently manifest as logarithmic terms proportional to $\ln(A)$ \cite{Kaul:2000iq, Carlip:2000nv, Adler:2001vs}, are a ubiquitous prediction across various quantum gravity approaches. Therefore, our thermodynamic analysis in this section is primarily aimed at capturing the leading-order thermal behavior driven by the GUP-modified horizon geometry, with the understanding that an exact thermodynamic mapping necessitates further sub-leading corrections to the entropy.

\subsection{The Fate of Singularity}

To study the behavior of the singularity and whether it is resolved, we
consider both the Kretschmann scalar $K$ and the affine parameters
of suitable null geodesics reaching $r=0$ from a larger $r$. Computing $K$ yields a complicated expression, so instead we use an approximate method to estimate $K$ near $r\to0$ in the equatorial plane $\theta=\frac{\pi}{2}$.
We consider the case of both the full metric and the slowly-rotating
limit.

\subsubsection{Full Rotating Metric}

To compute the dominant term of the Kretschmann scalar $K$ near $r\to0$
and at $\theta=\frac{\pi}{2}$, we need the leading terms of the Riemann
tensor components. We compute these components using Cartan’s second structure equation. This requires the spin connection, which in turn is obtained from diagonal dual tetrads.
The latter can be read off from the metric if we write it in a ``diagonalized
form''. 

To follow this procedure, we note that at $\theta=\frac{\pi}{2}$
and near $r\to0$ we have
\begin{align}
\rho^{2}\big|_{r\to0}= & \rho^{2}_{0}+\mathcal{O}\left(r^{8}\right),\\
\Delta\big|_{r\to0}= & a^{2}+\mathcal{O}\left(r^{2}\right),\\
\Sigma\big|_{r\to0}= & \rho^{2}_{0}\frac{r}{\sqrt{Q_{b}}}+\mathcal{O}\left(r^{3}\right),\\
\varLambda\big|_{r\to0}= & \rho^{2}_{0}\frac{r}{\sqrt{Q_{b}}}-a^{2}+\mathcal{O}\left(r^{2}\right).
\end{align}
where 
\begin{equation}
\rho^{2}_{0}=\left(\frac{Q_{c}R^{2}_{s}}{4}\right)^{\frac{1}{4}}.\label{eq:rho0-def}
\end{equation}
So the dominant terms of metric components in this approximation are
\begin{align}
g_{\bar{0}\bar{0}}\big|_{r\to0}\approx & -\frac{Q_{b}}{\rho^{2}_{0}}\left(1-\frac{R_{s}}{\sqrt{Q_{b}}}\right),\label{eq:g00-r0-theta-pi2}\\
g_{\bar{0}\bar{3}}\big|_{r\to0}\approx & -\frac{a\sqrt{Q_{b}}}{r},\label{eq:g03-r0-theta-pi2}\\
g_{\bar{1}\bar{1}}\big|_{r\to0}\approx & \frac{\rho^{2}_{0}}{a^{2}},\label{eq:g11-r0-theta-pi2}\\
g_{\bar{2}\bar{2}}\big|_{r\to0}\approx & \rho^{2}_{0},\label{eq:g22-r0-theta-pi2}\\
g_{\bar{3}\bar{3}}\big|_{r\to0}\approx & \rho^{2}_{0}+2a^{2}\frac{\sqrt{Q_{b}}}{r},\label{eq:g33-r0-theta-pi2}
\end{align}
For future reference, one can check that the inverse metric components'
dominant behaviors are
\begin{align}
g^{\bar{0}\bar{0}}\big|_{r\to0}\approx & \mathcal{O}\left(r\right),\\
g^{\bar{0}\bar{3}}\big|_{r\to0}\approx & \mathcal{O}\left(r\right),\\
g^{\bar{1}\bar{1}}\big|_{r\to0}\approx & \mathcal{O}\left(1\right),\\
g^{\bar{2}\bar{2}}\big|_{r\to0}\approx & \mathcal{O}\left(1\right),\\
g^{\bar{3}\bar{3}}\big|_{r\to0}\approx & \mathcal{O}\left(r^{2}\right)
\end{align}
We now use $g_{ab}=\eta_{IJ}e^{I}_{a}e^{J}_{b}$, where $I,\,J$ are
the $so(1,3)$ internal indices and assume diagonal dual tetrads to
find $e^{I}_{a}$ from 
\begin{equation}
ds^{2}=-\left(e^{0}\right)^{2}+\left(e^{1}\right)^{2}+\left(e^{2}\right)^{2}+\left(e^{3}\right)^{2}.\label{eq:ds-e2-diag}
\end{equation}
In this last equation, indices are associated to $so(1,3)$. To achieve
this, we need to diagonalize the line element. Since there are no
cross terms $drdx^{\mu},\,\mu\neq r$ and $d\theta dx^{\mu},\,\mu\neq\theta$
in the metric, we only need to diagonalize the $t-\phi$ part of the
metric which at the moment looks like 
\begin{equation}
ds^{2}_{t\phi}=g_{\bar{0}\bar{0}}dt^{2}+2g_{\bar{0}\bar{3}}dtd\phi+g_{\bar{3}\bar{3}}d\phi^{2}.
\end{equation}
Writing this metric as
\begin{equation}
ds^{2}_{t\phi}=g_{33}\left(d\phi^{2}+2\frac{g_{03}}{g_{33}}dtd\phi\right)+g_{00}dt^{2}
\end{equation}
and completing the square for the parenthesis term, we obtain the full
metric as
\begin{equation}
ds^{2}=-\left(\frac{g^{2}_{\bar{0}\bar{3}}}{g_{\bar{3}\bar{3}}}-g_{\bar{0}\bar{0}}\right)dt^{2}+g_{\bar{1}\bar{1}}dr^{2}+g_{\bar{2}\bar{2}}d\theta^{2}+g_{\bar{3}\bar{3}}\left(d\phi+\frac{g_{\bar{0}\bar{3}}}{g_{\bar{3}\bar{3}}}dt\right)^{2}.
\end{equation}
The ratio $-\frac{g_{03}}{g_{33}}$ is sometimes referred to as the
frame-dragging angular velocity of the spacetime. Comparing this form
of the metric with \eqref{eq:ds-e2-diag} and our components \eqref{eq:g00-r0-theta-pi2}-\eqref{eq:g33-r0-theta-pi2},
we can read off the dual tetrads as (we have suppressed the abstract
indices $a,\,b,\ldots$)
\begin{align}
e^{0}= & \sqrt{\frac{g^{2}_{\bar{0}\bar{3}}}{g_{\bar{3}\bar{3}}}-g_{\bar{0}\bar{0}}}dt\approx\left(\frac{\sqrt{Q_{b}}}{2r}\right)^{\frac{1}{2}}dt\\
e^{1}= & \sqrt{g_{\bar{1}\bar{1}}}dr\approx\frac{\rho_{0}}{a}dr\\
e^{2}= & \sqrt{g_{\bar{2}\bar{2}}}d\theta\approx\rho_{0}d\theta\\
e^{3}= & \sqrt{g_{\bar{3}\bar{3}}}\left(d\phi+\frac{g_{\bar{0}\bar{3}}}{g_{\bar{3}\bar{3}}}dt\right)\approx a\sqrt{\frac{2}{r}}Q^{\frac{1}{4}}_{b}\left(d\phi-\frac{1}{2a}dt\right)
\end{align}
where in computing $e^{0}$ and $e^{3}$ above, we have neglected
the $\mathcal{O}\left(1\right)$ term $g_{\bar{0}\bar{0}}$, and, in $g_{\bar{3}\bar{3}}$, the $\mathcal{O}\left(1\right)$ term $\rho^{2}_{0}$, keeping only $g_{\bar{3}\bar{3}}\approx2a^{2}\frac{\sqrt{Q_{b}}}{r}$.
We can invert the above equations and write down the coordinate basis
one forms as
\begin{align}
dt= & \left(\frac{\sqrt{Q_{b}}}{2r}\right)^{-\frac{1}{2}}e^{0},\label{eq:dt-e0}\\
dr= & \frac{a}{\rho_{0}}e^{1},\label{eq:dr-e1}\\
d\theta= & \frac{1}{\rho_{0}}e^{2},\label{eq:dtheta-e2}\\
d\phi= & \frac{1}{a}\left(Q_{b}\right)^{-\frac{1}{4}}\sqrt{\frac{r}{2}}\left(e^{3}+e^{0}\right).\label{eq:dphi-e0-e3}
\end{align}
In order to find the spin connection 1-form $\omega^{I}{}_{J}$ we
use the torsionless Cartan's first structure equation 
\begin{equation}
de^{I}+\omega^{I}{}_{J}\wedge e^{J}=0.\label{eq:Cartan's-1st}
\end{equation}
Thus, we need the exterior derivatives $de^{I}$. From \eqref{eq:dt-e0}-\eqref{eq:dphi-e0-e3}
we see that $de^{1}=0=de^{2}$, while the nonzero components
become
\begin{align}
de^{0}= & -\frac{1}{2}\left(\frac{\sqrt{Q_{b}}}{2r^{3}}\right)^{\frac{1}{2}}dr\wedge dt=\frac{a}{2r\rho_{0}}e^{0}\wedge e^{1},\\
de^{3}= & \sqrt{\frac{1}{2r^{3}}}Q^{\frac{1}{4}}_{b}\left(-a\,dr\wedge d\phi+\frac{1}{2}dr\wedge dt\right)=\frac{a}{2r\rho_{0}}e^{3}\wedge e^{1}.
\end{align}
Matching the above result with Eq. \eqref{eq:Cartan's-1st}, we obtain
\begin{align}
\omega^{0}{}_{1}=\omega^{1}{}_{0}= & -\frac{a}{2r\rho_{0}}e^{0},\label{eq:spin-conn-01}\\
\omega^{3}{}_{1}=-\omega^{1}{}_{3}= & -\frac{a}{2r\rho_{0}}e^{3}.\label{eq:spin-conn-31}
\end{align}
Next, we use the second Cartan's structure equation
\begin{equation}
R^{I}{}_{J}=d\omega^{I}{}_{J}+\omega^{I}{}_{K}\wedge\omega^{K}{}_{J},\label{eq:Cartan-2nd}
\end{equation}
to find the components of the curvature 2-form. The nonvanishing terms
are
\begin{align}
R^{0}{}_{1}= & -\frac{3}{4}\frac{a^{2}}{r^{2}\rho^{2}_{0}}e^{0}\wedge e^{1},\label{eq:R01}\\
R^{0}{}_{3}= & -\frac{1}{4}\frac{a^{2}}{r^{2}\rho^{2}_{0}}e^{0}\wedge e^{3},\label{eq:R03}\\
R^{1}{}_{3}= & -\frac{3}{4}\frac{a^{2}}{r^{2}\rho^{2}_{0}}e^{1}\wedge e^{3}.\label{eq:R13}
\end{align}
From this we can find the components of the Riemann tensor (since
the tetrads are diagonal) as
\begin{align}
R^{0}{}_{101}= & -\frac{3}{4}\frac{a^{2}}{r^{2}\rho^{2}_{0}},\\
R^{0}{}_{303}= & -\frac{1}{4}\frac{a^{2}}{r^{2}\rho^{2}_{0}},\\
R^{1}{}_{313}= & -\frac{3}{4}\frac{a^{2}}{r^{2}\rho^{2}_{0}}.
\end{align}
Finally, the Kretschmann scalar is obtained as
\begin{align}
K= & R^{IJKL}R_{IJKL}\nonumber \\
= & 4\left(R^{0101}R_{0101}+R^{0303}R_{0303}+R^{1313}R_{1313}\right)\nonumber \\
= & \frac{19}{2}\frac{a^{4}}{\sqrt{Q_{c}}R_{s}}\frac{1}{r^{4}},\label{eq:Kretsch-full}
\end{align}
where we have used \eqref{eq:rho0-def}. This is a softer singularity
compared to the classical Kerr black hole that goes as $r^{-6}$,
but it is still divergent.

Let us now study the affine parameter distance to $r=0$. Considering
a null geodesic $ds^{2}=0$ with $\theta=\frac{\pi}{2}$, from the
line element we obtain
\begin{equation}
0=g_{\bar{0}\bar{0}}\dot{t}^{2}+2g_{\bar{0}\bar{3}}\dot{t}\dot{\phi}+g_{\bar{3}\bar{3}}\dot{\phi}^{2}+g_{\bar{1}\bar{1}}\dot{r}^{2},
\end{equation}
where the dots denote derivatives with respect to the affine parameter
$\lambda$. The two Killing vector fields $\mathcal{K}^{\bar{\mu}}=\partial^{\bar{\mu}}_{\bar{0}}$
and $\mathcal{R}^{\bar{\mu}}=\partial^{\bar{\mu}}_{\bar{3}}$ of this
spacetime lead to the invariant energy $E$ and angular momentum $L$
\begin{align}
E= & -g_{\bar{\mu}\bar{\nu}}\mathcal{K}^{\mu}\dot{x^{\nu}}=-g_{\bar{0}\bar{0}}\dot{x^{0}}-g_{\bar{0}\bar{3}}\dot{x^{3}}=-g_{\bar{0}\bar{0}}\dot{t}-g_{\bar{0}\bar{3}}\dot{\phi}\\
L= & g_{\bar{\mu}\bar{\nu}}\mathcal{R}^{\mu}\dot{x^{\nu}}=g_{\bar{0}\bar{3}}\dot{x^{0}}+g_{\bar{3}\bar{3}}\dot{x^{3}}=g_{\bar{0}\bar{3}}\dot{t}+g_{\bar{3}\bar{3}}\dot{\phi}
\end{align}
of the infalling object along the geodesic. The solution to these
are
\begin{align}
\dot{t}= & -\frac{-Eg_{\bar{3}\bar{3}}-Lg_{\bar{0}\bar{3}}}{g^{2}_{\bar{0}\bar{3}}-g_{\bar{0}\bar{0}}g_{\bar{3}\bar{3}}}\approx-\frac{-\mathcal{O}\left(r^{-1}\right)E-\mathcal{O}\left(r^{-1}\right)L}{\mathcal{O}\left(r^{-2}\right)-\mathcal{O}\left(r^{-1}\right)}\approx\mathcal{O}\left(r\right),\label{eq:t-dot-O}\\
\dot{\phi}= & -\frac{Eg_{\bar{0}\bar{3}}+Lg_{\bar{0}\bar{0}}}{g^{2}_{\bar{0}\bar{3}}-g_{\bar{0}\bar{0}}g_{\bar{3}\bar{3}}}\approx-\frac{\mathcal{O}\left(r^{-1}\right)E-\mathcal{O}\left(1\right)L}{\mathcal{O}\left(r^{-2}\right)-\mathcal{O}\left(r^{-1}\right)}\approx\mathcal{O}\left(r\right).\label{eq:phi-dot-O}
\end{align}
Substituting these into the null condition $ds^{2}=0$, we obtain
\begin{align}
0= & g_{\bar{0}\bar{0}}\dot{t}^{2}+2g_{\bar{0}\bar{3}}\dot{t}\dot{\phi}+g_{\bar{3}\bar{3}}\dot{\phi}^{2}+g_{\bar{1}\bar{1}}\dot{r}^{2}\nonumber \\
\approx & \mathcal{O}\left(1\right)\mathcal{O}\left(r^{2}\right)+\mathcal{O}\left(r^{-1}\right)\mathcal{O}\left(r^{2}\right)+\mathcal{O}\left(r^{-1}\right)\mathcal{O}\left(r^{2}\right)+g_{\bar{1}\bar{1}}\dot{r}^{2}\nonumber \\
\approx & \mathcal{O}\left(r^{2}\right)+\mathcal{O}\left(r\right)+\mathcal{O}\left(r\right)+g_{\bar{1}\bar{1}}\dot{r}^{2}.
\end{align}
For $r\to0$, the leading order behavior is therefore
\begin{equation}
g_{\bar{1}\bar{1}}\dot{r}^{2}\approx\mathcal{O}\left(r\right),
\end{equation}
and since $g_{\bar{1}\bar{1}}\approx\mathcal{O}(1)$ we get
\begin{equation}
\dot{r}^{2}\sim r.
\end{equation}
This leads to 
\begin{equation}
\frac{dr}{d\lambda}\sim r^{\frac{1}{2}},
\end{equation}
and as a result
\begin{equation}
\Delta\lambda\sim\int^{0}_{r_{0}}r^{-\frac{1}{2}}dr\sim\sqrt{r_{0}}.
\end{equation}
Hence, $K(r\to0)\to\infty$ while $r=0$ can be reached in finite
affine parameter. This implies that the ring singularity in this model
is still present at $r=0,\,\theta=\frac{\pi}{2}$. This seems to be
a property of some of the rotating black hole models that are derived
using the NJ  algorithm. In these models, although the static regular black hole seed can be nonsingular, applying the NJ algorithm does not generically preserve all desirable properties and can reintroduces Kerr-like pathologies (including singular behavior, energy-condition violations, or naked-singularity branches), depending on the construction \cite{HansenYunes2013,KamenshchikPetriakova2023,ShaoChenChen2021,NevesSaa2014}. However,
we will show in the next section that the slowly-rotating version
of this metric avoids this singularity.

\subsubsection{Slowly-Rotating Metric}

As previously discussed, the slowly-rotating limit of this metric is
the $a^{2}\to0$ limit of the full metric. It is has a much simpler
form and can be written as
\begin{align}
g^{\text{(slow)}}_{\bar{0}\bar{0}}= & g_{00}(r),\label{eq:GUP-slow-rot-00}\\
g^{\text{(slow)}}_{\bar{0}\bar{3}}= & -a\sin^{2}\left(\theta\right)\left[\left(1+\frac{Q_{b}}{r^{2}}\right)^{\frac{1}{2}}+g_{00}(r)\right],\label{eq:GUP-slow-rot-03}\\
g^{\text{(slow)}}_{\bar{1}\bar{1}}= & g_{11}(r),\label{eq:GUP-slow-rot-11}\\
g^{\text{(slow)}}_{\bar{2}\bar{2}}= & g_{22}(r),\label{eq:GUP-slow-rot-22}\\
g^{\text{(slow)}}_{\bar{3}\bar{3}}= & g_{33}(r).\label{eq:GUP-slow-rot-33}
\end{align}
It is seen that the only difference between the slowly-rotating GUP
metric \eqref{eq:GUP-slow-rot-00}-\eqref{eq:GUP-slow-rot-33} and
the static GUP metric \eqref{eq:GUP-non-rot-00}-\eqref{eq:GUP-non-rot-33}
is the presence of $g^{\text{(slow)}}_{\bar{0}\bar{3}}$ component,
while all other metric components are the same.

A similar computation to the previous section reveals that the Kretschmann
scalar in this case is finite and becomes $K=\frac{8}{R_{s}\sqrt{Q_{c}}}$.
We can also show that it takes infinite affine parameter to reach
$r=0$. In this case, Eqs. \eqref{eq:t-dot-O} and \eqref{eq:phi-dot-O}
reduce to
\begin{align}
\dot{t}= & -\frac{-Eg^{\text{(slow)}}_{\bar{3}\bar{3}}-Lg^{\text{(slow)}}_{\bar{0}\bar{3}}}{\left[g^{\text{(slow)}}_{\bar{0}\bar{3}}\right]^{2}-g^{\text{(slow)}}_{\bar{0}\bar{0}}g^{\text{(slow)}}_{\bar{3}\bar{3}}}\approx-\frac{-\mathcal{O}\left(1\right)E-\mathcal{O}\left(r^{-1}\right)L}{\mathcal{O}\left(r^{-2}\right)-\mathcal{O}\left(1\right)}\approx\mathcal{O}\left(r\right),\\
\dot{\phi}= & -\frac{Eg^{\text{(slow)}}_{\bar{0}\bar{3}}+Lg^{\text{(slow)}}_{\bar{0}\bar{0}}}{\left[g^{\text{(slow)}}_{\bar{0}\bar{3}}\right]^{2}-g^{\text{(slow)}}_{\bar{0}\bar{0}}g^{\text{(slow)}}_{\bar{3}\bar{3}}}\approx-\frac{\mathcal{O}\left(r^{-1}\right)E-\mathcal{O}\left(1\right)L}{\mathcal{O}\left(r^{-2}\right)-\mathcal{O}\left(r^{-1}\right)}\approx\mathcal{O}\left(r\right).
\end{align}
Replacing these back into the null condition $ds^{2}=0$, we obtain
\begin{align}
0= & g^{\text{(slow)}}_{\bar{0}\bar{0}}\dot{t}^{2}+2g^{\text{(slow)}}_{\bar{0}\bar{3}}\dot{t}\dot{\phi}+g^{\text{(slow)}}_{\bar{3}\bar{3}}\dot{\phi}^{2}+g^{\text{(slow)}}_{\bar{1}\bar{1}}\dot{r}^{2}\nonumber \\
\approx & \mathcal{O}\left(1\right)\mathcal{O}\left(r^{2}\right)+\mathcal{O}\left(r^{-1}\right)\mathcal{O}\left(r^{2}\right)+\mathcal{O}\left(1\right)\mathcal{O}\left(r^{2}\right)+g_{\bar{1}\bar{1}}\dot{r}^{2}\nonumber \\
\approx & \mathcal{O}\left(r^{2}\right)+\mathcal{O}\left(r\right)+\mathcal{O}\left(r^{2}\right)+g_{\bar{1}\bar{1}}\dot{r}^{2}.
\end{align}
For $r\to0$, the leading order behavior is still
\begin{equation}
g^{\text{(slow)}}_{\bar{1}\bar{1}}\dot{r}^{2}\approx\mathcal{O}\left(r\right).
\end{equation}
However, in this case we have $g^{\text{(slow)}}_{\bar{1}\bar{1}}\approx\mathcal{O}\left(r^{-2}\right)$
and hence
\begin{equation}
\dot{r}^{2}\sim r^{3}.
\end{equation}
This leads to 
\begin{equation}
\frac{dr}{d\lambda}\sim r^{\frac{3}{2}},
\end{equation}
and as a result
\begin{equation}
\Delta\lambda\sim\int^{0}_{r_{0}}r^{-\frac{3}{2}}dr\sim\left[\frac{1}{\sqrt{r}}\right]^{0}_{r_{0}}\sim\infty.
\end{equation}
Thus the singularity is actually resolved in the slowly-rotating metric.
A physical interpretation of the fact that the singularity is resolved
in the slowly-rotating case but not in the fully rotating scenario
is that the slow-rotation limit effectively creates a logarithmic
throat geometry, which explains both the infinite affine parameter
and the very strong curvature divergence we computed earlier. To see
the existence of this wormhole-like geometry, we check the proper
radial distance and the area of the 2-spheres as $r\to0$. If the
proper distance diverges while the angular area approaches a finite
non-zero value, the geometry develops a cylindrical (or infinite)
throat.

For the proper radial distance we have
\begin{align}
\ell\left(r\right)= & \int^{0}_{r_{0}}\sqrt{g^{\text{(slow)}}_{\bar{1}\bar{1}}}dr\nonumber \\
\sim & \int^{0}_{r_{0}}\sqrt{r^{-2}}dr\nonumber \\
\sim & \left[\ln\left(r\right)\right]^{0}_{r_{0}}\nonumber \\
\sim & \infty.
\end{align}
So the first condition
\begin{equation}
\ell\left(r\to0\right)\to\infty,
\end{equation}
is satisfied. The second condition can be easily verified by considering the area of the 2-spheres 
\begin{equation}
A=4\pi\sqrt{g^{\text{(slow)}}_{\bar{2}\bar{2}}g^{\text{(slow)}}_{\bar{3}\bar{3}}}\approx\mathcal{O}\left(1\right).
\end{equation}
Since the angular area approaches a finite, nonzero value as $r\to0$ the second condition is also satisfied.
Furthermore, a coordinate transformation
\begin{equation}
r=e^{-\ell}
\end{equation}
near $r\to0$ casts the $rr,\,\theta\theta,$ and $\phi\phi$ part
of the metric into
\begin{equation}
ds^{2}\approx d\ell^{2}+R^{2}_{0}d\Omega^{2},
\end{equation}
which is the metric of a cylindrical throat.

In hindsight, one can argue that the slowly-rotating metric is closer
to the static metric in which the singularity is resolved. So it seems
that the NJ  algorithm introduces the singularity back into
the full rotating metric by structurally shearing the regularized
core at the equator.

\begin{figure}[tp]
\centering \includegraphics[width=0.8\linewidth]
{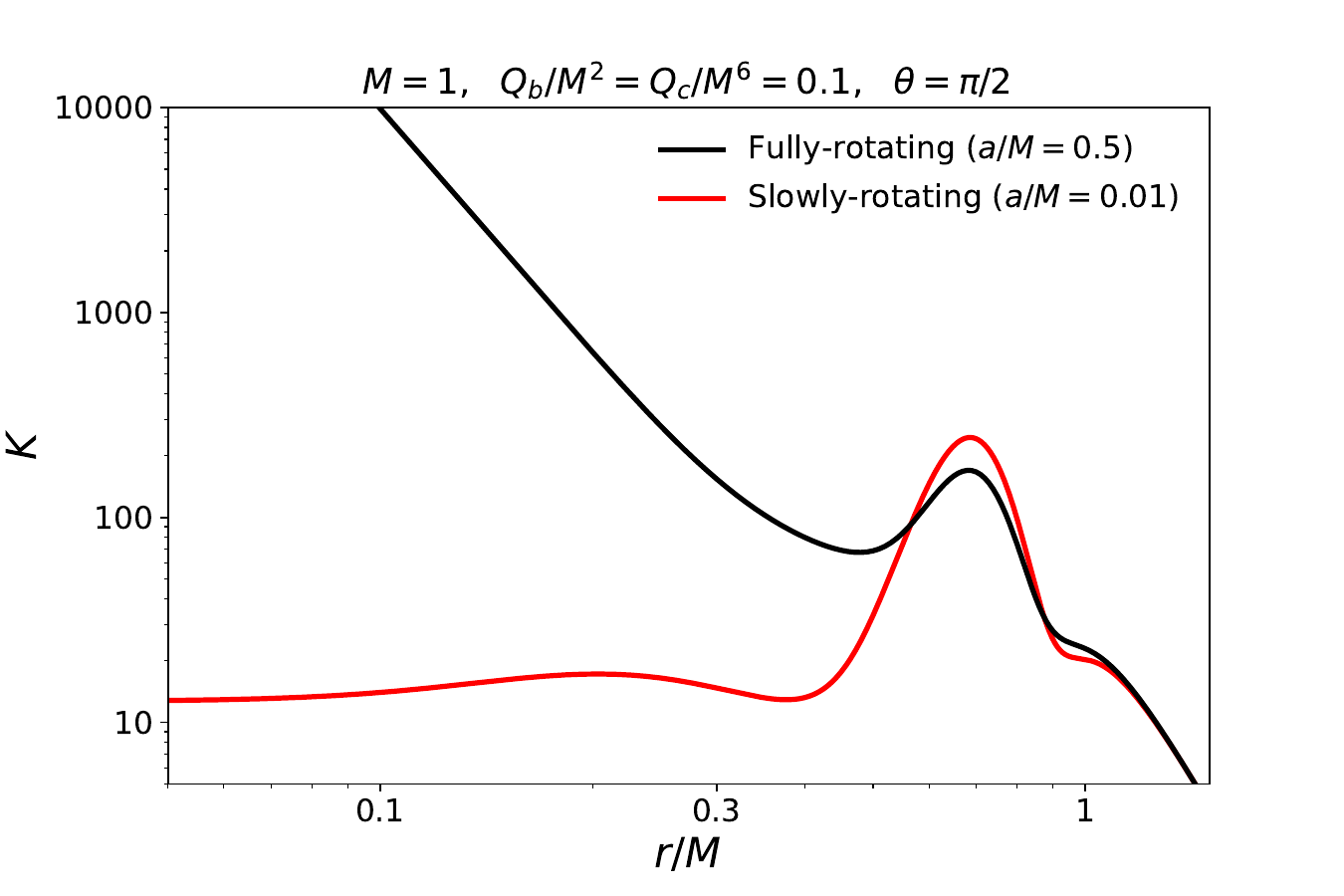}
\caption{Plot of the Kretschmann scalar as a function of the radial coordinate for the fully rotating (with $\frac{a}{M} = 0.5$) and slowly-rotating (with $\frac{a}{M} = 0.01$) black holes, with the values of the other parameters given on the top of the plot.}\label{kscalar}
\end{figure}

Figure \ref{kscalar} shows the full numerical expression of $K$ as a function of the radial coordinate for the fully rotating and slowly-rotating black holes. This plot shows the difference in behavior of $K$, namely that $K$ for the fully rotating black hole diverges for small $r$ whereas $K$ for the slowly-rotating black hole remains finite. Due to computational constraints, these plots are cut off at an arbitrary value of $r$. However, the limit of $K$ as $r \rightarrow 0$ computed numerically  shows that the fully rotating $K$ goes to $\infty$ in this limit, while the slowly-rotating $K$ remains finite.

\section{Shadow and Phenomenology\label{Sec:shadow}}

\subsection{The Shape of the Shadow}

With the rotating metric in hand, we are now interested in comparing
the shadow of this black hole with observations from the Event Horizon Telescope (EHT) of the black hole at the center
of our galaxy, Sgr A{*}, and the one at the heart of the M87 galaxy, known as M87*.

Rotating black holes do not have a photon sphere, instead they have
a photon region. There is a counter-rotating circular orbit in the
equatorial plane of the black hole at $r^{\text{ph}}_{-}$ and a co-rotating
circular orbit in the equatorial plane of the black hole at $r^{\text{ph}}_{+}$.
The light rays in between these two orbits are non-planar and spherical
(move on a spherical shell). We have $r^{\text{ph}}_{-}>r^{\text{ph}}_{+}$
and the region between these two which includes the non-planar spherical
light orbits is called the photon region. Imagine a black hole rotating
counter clockwise with respect to us, i.e., such that its left side
is coming towards us and its right side going away from us. Then the
light from the left side that we see coming towards us is emanating
from $r^{\text{ph}}_{+}$ and the light from the right side is coming
from $r^{\text{ph}}_{-}$. The shadow is the boundary of past oriented
light rays emanated from the observer that asymptotically end up on
unstable spherical orbits around the black hole. Due to the fact that
this boundary on the left and right is associated to rays that do
not have the same orbit radius as mentioned above, the shadow of a
rotating black hole is not a circle; it is oblique \cite{Perlick:2021aok}. The spherical null geodesic that is the limiting curve
for these past oriented light ray that emanated from the observer
must have the same constants of motion $L,\,E$, i.e., angular momentum
and energy respectively, which are determined by its radius $r_{p}$. 

The contour of the rotating black hole shadow can be obtained by finding
the unstable circular orbit of null geodesics, as outlined in Refs.
\cite{Perlick:2021aok,Islam:2022wck,Devi:2021ctm}. The apparent
shape of the shadow as seen by an observer located at asymptotically
flat infinity, with an inclination angle $\theta_{0}$ is given by
the celestial coordinates $\alpha,\beta$ given by~\cite{Perlick:2021aok}
\begin{align}
\alpha\left(r_{p}\right)= & -\xi\left(r_{p}\right)\csc\left(\theta_{0}\right),\\
\beta\left(r_{p}\right)= & \pm\sqrt{\eta\left(r_{p}\right)+a^{2}\cos^{2}\left(\theta_{0}\right)-\xi\left(r_{p}\right)^{2}\cot^{2}\left(\theta_{0}\right)}.
\end{align}
These are computed using the impact parameters for unstable circular
photon orbits,  given by
\begin{align}
\xi\left(r_{p}\right)= & \frac{\Upsilon\left(r_{p}\right)\Delta^{\prime}\left(r_{p}\right)-2\Delta\left(r_{p}\right)\Upsilon^{\prime}\left(r_{p}\right)}{a\Delta^{\prime}\left(r_{\text{ps}}\right)}\\
\eta\left(r_{p}\right)= & \frac{4a^{2}\Upsilon^{\prime}\left(r_{p}\right)^{2}\Delta\left(r_{p}\right)-\left[\left(\Upsilon\left(r_{p}\right)-a^{2}\right)\Delta^{\prime}\left(r_{p}\right)-2\Upsilon^{\prime}\left(r_{p}\right)\Delta\left(r_{p}\right)\right]^{2}}{a^{2}\Delta^{\prime}\left(r_{p}\right)}
\end{align}

\noindent where the prime denotes the derivative with respect to the
radial coordinate $r$, and $\Upsilon(r)$ is given by
\begin{equation}
\Upsilon=\Sigma+a^{2}\sin^{2}\left(\theta\right).
\end{equation}
Furthermore, the first impact parameters $\xi=\frac{L}{E}$, is the
photon's perpendicular distance from the rotation axis of the black
hole as seen by a distant observer, and the second impact parameter,
$\eta=\frac{\mathcal{K}}{E}$, measures the motion out of the equatorial
plane. Here $\mathcal{K}$ is the Carter constant. 

\begin{figure}[tp]
\centering \includegraphics[width=0.6\linewidth]{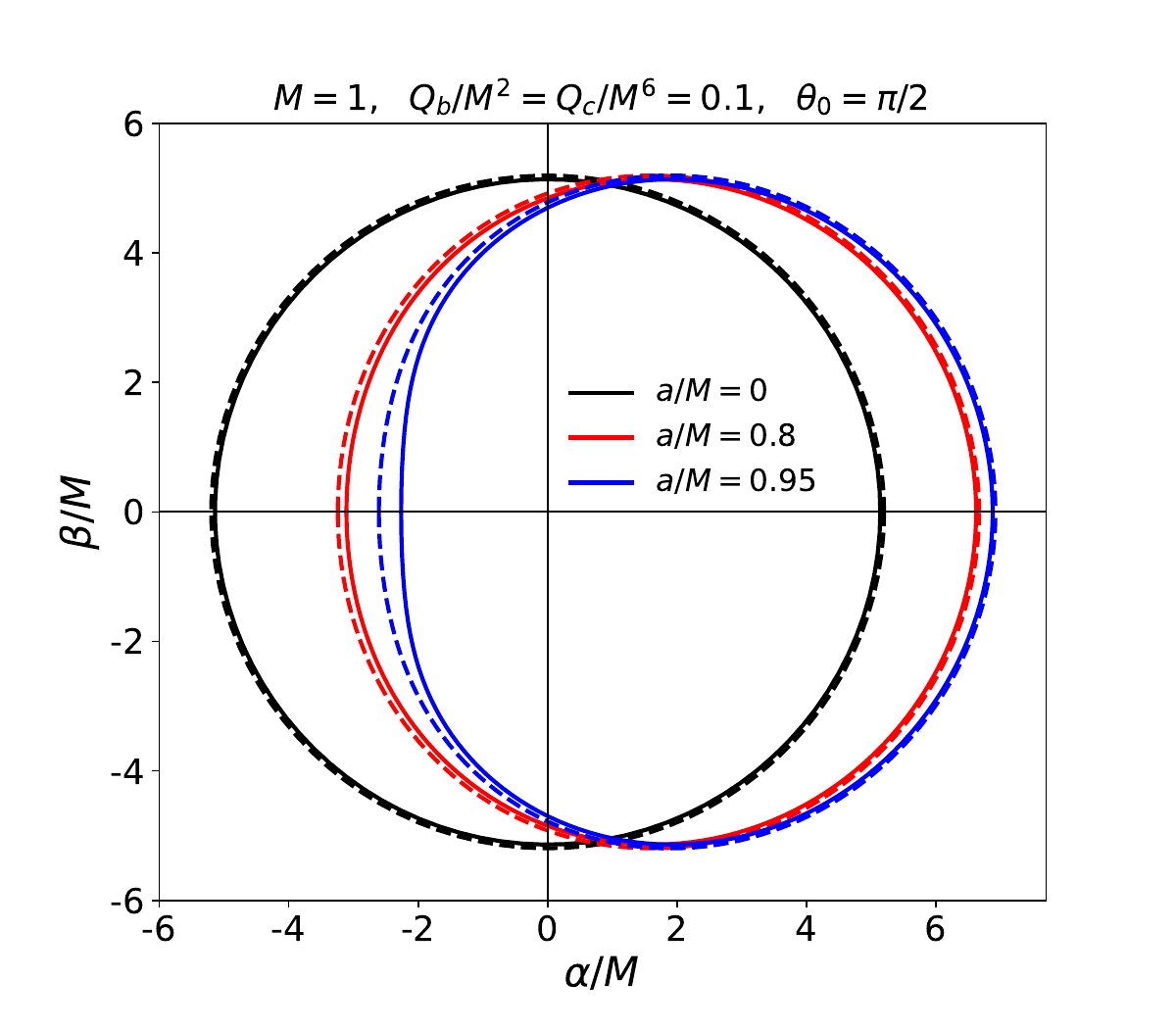}
\caption{Shadow contours in celestial coordinates for three values of the spin
parameter, $\frac{a}{M}=0$ (black lines), $\frac{a}{M}=0.8$ (red
lines) and $\frac{a}{M}=0.95$ (blue lines). Solid lines correspond
to the shadow for this rotating GUP black hole with quantum parameters
$\frac{Q_{b}}{M^{2}}=0.1=\frac{Q_{c}}{M^{6}}$, dashed lines correspond
to the respective Kerr black hole $\left(\frac{Q_{b}}{M^{2}}=0=\frac{Q_{c}}{M^{6}}\right)$.
The black hole mass is $M=1$ and the inclination angle is $\theta_{0}=\frac{\pi}{2}$.}
\label{shadow_contours} 
\end{figure}

Fig.~\ref{shadow_contours} shows the black hole shadow contours in
celestial coordinates for three values of the spin parameter, for
this GUP rotating black hole with $\frac{Q_{b}}{M^{2}}=0.1=\frac{Q_{c}}{M^{6}}$
(solid lines) and the corresponding classical Kerr black hole (dashed
lines). The impact of the quantum parameter $Q_{b}$ becomes larger
as the spin of the black hole increases, with the largest deviation
between the quantum and classical shadows appearing on the distorted
side of the shadow.

\subsection{Bounds on Quantum Parameter $Q_{b}$ From Sgr A* And M87*}

We now make use of shadow observables to constrain the quantum parameter
$Q_{b}$. The other parameter $Q_{c}$ has a much smaller effect on
the shadow and contributes mostly to the quantum effects in the interior
of the black hole, particularly near the singularity. The shadows
of the Sgr A* and M87* black holes have been observed by the EHT
\cite{EventHorizonTelescope:2019dse,EventHorizonTelescope:2022wkp}.
These observations provide measurements of the angular shadow diameter
$d_{\text{sh}}$ and Schwarzschild shadow deviation $\delta$ defined
as
\begin{align}
d_{{\rm sh}}= & \frac{2}{D}\sqrt{\frac{A}{\pi}},\\
\delta= & \frac{d_{\text{sh}}}{d_{\text{Sch}}}-1,
\end{align}
where $D$ is the distance from the observer to the black
hole, $d_{\text{Sch}}$ is the corresponding Schwarzschild angular
shadow diameter of the black hole (i.e., for $a=0$), and $A$ is the
shadow area defined as
\begin{equation}
A=2\int^{r^{+}_{p}}_{r^{-}_{p}}\beta\left(r_{p}\right)\frac{d\alpha\left(r_{p}\right)}{dr_{p}}dr_{p}.
\end{equation}
Here $r^{-}_{p}$ and $r^{+}_{p}$ are the minimal and maximal values
of $r_{p}$, respectively~\cite{Perlick:2021aok}.

\begin{figure}[tp]
\centering \includegraphics[width=1.0\linewidth]{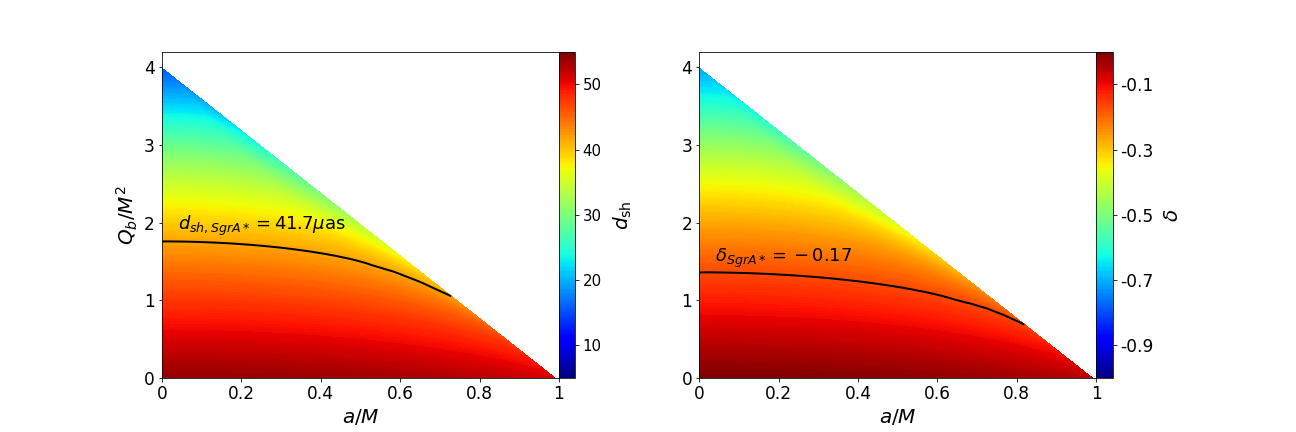}
\caption{Angular shadow diameter $d_{{\rm sh}}$ (left plot) and Schwarzschild
shadow deviation $\delta$ (right plot) for a range of quantum parameter
$Q_{b}$ values and spin parameter $a$ values for the Sgr A{*} black
hole, which has a mass, distance and inclination angle as given in
the text. The solid black lines indicate the lower bounds on the values
of $d_{{\rm sh}}$ and $\delta$, respectively, for Sgr A{*} as measured
from EHT observations. The white region shows the values of $Q_{b}$
and $a$ for which the black hole has no horizons.}
\label{constraints_sgr} 
\end{figure}

For Sgr A$^{*}$, EHT measures the angular shadow diameter to be $d_{\text{sh, Sgr\ensuremath{A^{*}}}}=48.7\pm7.0\hspace{0.2cm}\mu\textrm{as}$,
using solely EHT data \cite{EventHorizonTelescope:2022wkp}. They
obtain two measurements for the Schwarzschild shadow deviation using
two different distance measurements as priors. We will consider the
Schwarzschild shadow deviation measurement of $\delta_{{\rm SgrA^{*}}}=-0.08\pm0.09$,
which utilizes the most recent distant measurement of $D_{{\rm SgrA^{*}}}=8.277\pm0.033$
kpc, as obtained from Very Large Telescope Interferometer observations
of stellar orbits \cite{GRAVITY:2021xju}. Both of the above shadow
observables are given at the 1$\sigma$ confidence level. For the
mass of Sgr A{*}, we use the value as measured by \cite{GRAVITY:2021xju},
given by $M_{{\rm SgrA^{*}}}=(4.297\pm0.013)\times10^{6}\,\text{M}_{\odot}$. The inclination angle of Sgr A$^{*}$ has been constrained to lie
in the range of $40$ - $60$ deg. We consider the median value of
$\theta_{0}=50$ deg.

For M87*, EHT measures the angular shadow diameter to be in the range
of $39\,\mu\textrm{as}\lesssim d_{{\rm sh,M87*}}\lesssim45\,\mu\textrm{as}$
and the Schwarzschild shadow deviation to be $\delta_{{\rm M87*}}=-0.01\pm0.17$,
at the 1$\sigma$ confidence level \cite{EventHorizonTelescope:2019dse,EventHorizonTelescope:2019pgp}.
These observations are based on a distance of $D_{{\rm M87*}}=16.8$
Mpc and an estimated mass of $M_{{\rm M87*}}=(6.5\pm0.7)\times10^{9}\,\text{M}_{\odot}$. The inclination angle of M87* has been constrained to $\theta_{0}\sim17$
deg, which we use in our analysis.

\begin{figure}[tp]
\centering \includegraphics[width=1\linewidth]{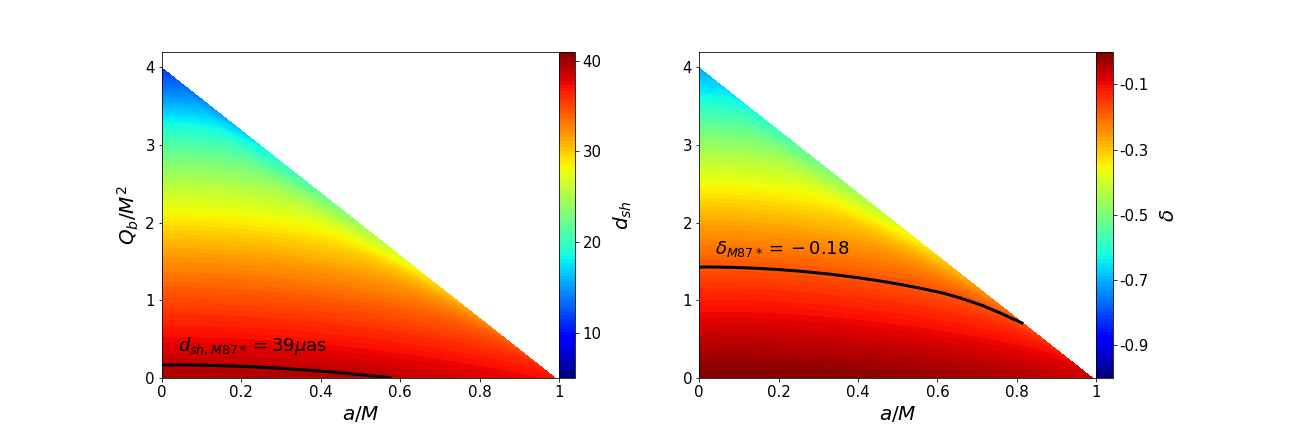}
\caption{Same as Fig.~\ref{constraints_sgr} but for the M87* black hole, which
has a mass, distance and inclination angle as given in the text. The
solid black lines indicate the lower bounds on the values of $d_{{\rm sh}}$
and $\delta$, respectively, for M87* as measured from EHT observations.}
\label{constraints_m87} 
\end{figure}

Figs.~\ref{constraints_sgr} and \ref{constraints_m87} show the angular
shadow diameter $d_{{\rm sh}}$ (left plot) and Schwarzschild shadow
deviation $\delta$ (right plot) for a range of quantum parameter
values $Q_{b}$, and spin parameter values $a$ the Sgr A$^{*}$
(Fig.~\ref{constraints_sgr}) and M87* (Fig.~\ref{constraints_m87}),
respectively. The solid black lines indicate the lower bounds on the
values of $d_{{\rm sh}}$ and $\delta$, for each respective black
hole. Any $(a,Q_{b})$ point that lies above the solid black lines
is ruled out as a result of these observations. The strongest constraint
comes from the angular shadow diameter of M87* which requires $\frac{Q_{b}}{M^{2}}\lesssim (0.001 \sim0.2)$ if M87* has a spin of $\frac{a}{M} < 0.6$. If M87* has a spin of $\frac{a}{M} > 0.6$, then $\frac{Q_b}{M^2}$ must be zero in order to be in agreement with EHT observations.

An interesting observation from Fig.~\ref{constraints_m87} is that
if this model is correct and if the numerical values of the observations
are accurate, then this quantum black hole cannot rotate faster than
$\frac{a}{M}\approx0.6$. If an observed angular momentum of this
black hole exceeds this value, it provides grounds for refuting this
model. 

\section{Conclusion\label{sec:Conclusion}}

In this work we have derived the metric for a rotating improved GUP
black hole by applying the modified NJ algorithm to a recent
metric of a static improved GUP black hole. Both these static and
rotating models have two quantum parameters $Q_{b}$ and $Q_{c}$.
The former is the main contributor to quantum gravity effects near
the horizon and outside of the black hole while the latter contributes
to the interior modification and quantum effects near $r=0$. 

The result of application of the NJ algorithm is a spinning
black hole that still has a maximum of two horizons, while the quantum
parameter $Q_{b}$ expands the inner horizon and shrinks the outer
one. In the extremal case the radius of the only horizon is smaller
than its classical counterpart and quantum effects allow a naked singularity for values of $\frac{a}{M}$ smaller than
one.

Although the singularity was resolved in the static case, the rotation
introduces it back into this rotating model. This is a rather well-known
side effect of the Newman-Janis algorithm. The slowly-rotating version
of this black hole, however, is singularity free, both due to the
finiteness of the Kretschmann scalar and that the $r=0,\,\theta=\frac{\pi}{2}$
region lies at infinite affine parameter.

We also compute the temperature and entropy of this black hole and
show that they both are smaller compared to the 
classical case due to quantum effects associated with $Q_{b}$.

An important part of this investigation is comparing the model with
observational data from the EHT. We have derived the shadow of this
black hole and by comparing it to the EHT, we set strong 
bounds on the quantum parameter $Q_{b}$, i.e., $\frac{Q_{b}}{M^{2}}\lesssim (0.001 \sim0.2)$ if M87* has a spin of $\frac{a}{M} < 0.6$. Furthermore, we show that if our
model is valid and if the EHT data is accurate, the M87* black hole cannot
rotate faster than $\frac{a}{M}\approx0.6$. However, the data is
still at the $1\sigma$ confidence level.


\acknowledgments{
The authors acknowledge the support of the Natural Sciences and Engineering Research Council of Canada (NSERC). We also thank Douglas Gingrich for valuable discussions.
}







\appendix


\section{Estimating the Kretschmann of the full metric}

We have seen that the dominant order of the metric, its
derivatives, and its inverse are
\begin{align*}
g\approx & \mathcal{O}\left(r^{-1}\right), & \partial_{r}g\approx & \mathcal{O}\left(r^{-2}\right) & g^{-1}\approx & \mathcal{O}\left(1\right).
\end{align*}
Consequently the Christoffel symbols behave as
\begin{equation}
\Gamma\approx g^{-1}\partial_{r}g\approx\mathcal{O}\left(1\right)\mathcal{O}\left(r^{-2}\right)\approx\mathcal{O}\left(r^{-2}\right),
\end{equation}
This is also because the dominant terms in $\Gamma$ is $g^{11}$
(of $\mathcal{O}(1)$; also $g^{22}$ does not matter since $\partial_{2}g$
vanishes for $\theta=\pi/2$), which is contracted by the dominant
terms $\partial_{1}g_{03}$ and $\partial_{1}g_{03}$ (of $\mathcal{O}\left(r^{-2}\right)$).
The Riemann tensor as
\begin{equation}
R\approx\partial_{r}\Gamma+\Gamma^{2}\approx\mathcal{O}\left(r^{-3}\right),
\end{equation}
since as mentioned above, the dominant terms of $\Gamma$ are $\Gamma^{1}_{33}$
and $\Gamma^{1}_{03}$ but they should be contracted in the Riemann
tensor as $\Gamma^{\rho}_{\mu\lambda}\Gamma^{\lambda}_{\nu\sigma}$.
But any $\Gamma$ with a lower index of $1$ (e.g., $\Gamma^{\rho}_{\mu1}$
) is at most $\mathcal{O}\left(r^{-1}\right)$, so $\Gamma^{2}$ goes
as $\mathcal{O}\left(r^{-3}\right)$ and not $\mathcal{O}\left(r^{-4}\right)$.

To check the order of the Kretschmann scalar, we note that the severe
divergences in the Riemann tensor come from the $t$ and $\phi$ indices
(indices $0$ and $3$). To raise these indices when calculating $K=R^{\alpha\beta\gamma\delta}R_{\alpha\beta\gamma\delta}$,
we are forced to use $g^{00}\approx\mathcal{O}(r),\,g^{03}\approx\mathcal{O}(r)$,
and $g^{33}\approx\mathcal{O}\left(r^{2}\right)$. Contracting $\mathcal{O}\left(r^{-3}\right)$
of the Riemann tensor with the dominant one of them, say $g^{00}$,
yields $\mathcal{O}\left(r^{-2}\right)$ and thus
\begin{equation}
K\approx\left(g^{-1}\right)^{4}R^{2}\approx\mathcal{O}\left(r^{-4}\right).
\end{equation}
This is the same result we obtained before using the diagonal tetrad
framework.


\bibliographystyle{jhep}





\bibliography{mainbib}{}

@article{Azreg-Ainou:2014aqa,
    author = {Azreg-A{\"\i}nou, Mustapha},
    title = "{From static to rotating to conformal static solutions: Rotating imperfect fluid wormholes with(out) electric or magnetic field}",
    eprint = "1401.4292",
    archivePrefix = "arXiv",
    primaryClass = "gr-qc",
    doi = "10.1140/epjc/s10052-014-2865-8",
    journal = "Eur. Phys. J. C",
    volume = "74",
    number = "5",
    pages = "2865",
    year = "2014"
}

@article{Azreg-Ainou:2014pra,
    author = {Azreg-A{\"\i}nou, Mustapha},
    title = "{Generating rotating regular black hole solutions without complexification}",
    eprint = "1405.2569",
    archivePrefix = "arXiv",
    primaryClass = "gr-qc",
    doi = "10.1103/PhysRevD.90.064041",
    journal = "Phys. Rev. D",
    volume = "90",
    number = "6",
    pages = "064041",
    year = "2014"
}

@article{Azreg-Ainou:2014nra,
    author = "Azreg-Ainou, Mustapha",
    title = "{Regular and conformal regular cores for static and rotating solutions}",
    eprint = "1401.0787",
    archivePrefix = "arXiv",
    primaryClass = "gr-qc",
    doi = "10.1016/j.physletb.2014.01.041",
    journal = "Phys. Lett. B",
    volume = "730",
    pages = "95--98",
    year = "2014"
}

@article{EventHorizonTelescope:2019pgp,
    author = "Akiyama, Kazunori and others",
    collaboration = "Event Horizon Telescope",
    title = "{First M87 Event Horizon Telescope Results. V. Physical Origin of the Asymmetric Ring}",
    eprint = "1906.11242",
    archivePrefix = "arXiv",
    primaryClass = "astro-ph.GA",
    doi = "10.3847/2041-8213/ab0f43",
    journal = "Astrophys. J. Lett.",
    volume = "875",
    number = "1",
    pages = "L5",
    year = "2019"
}

@article{EventHorizonTelescope:2022wkp,
    author = "Akiyama, Kazunori and others",
    collaboration = "Event Horizon Telescope",
    title = "{First Sagittarius A* Event Horizon Telescope Results. I. The Shadow of the Supermassive Black Hole in the Center of the Milky Way}",
    eprint = "2311.08680",
    archivePrefix = "arXiv",
    primaryClass = "astro-ph.HE",
    doi = "10.3847/2041-8213/ac6674",
    journal = "Astrophys. J. Lett.",
    volume = "930",
    number = "2",
    pages = "L12",
    year = "2022"
}

@article{Anacleto:2021qoe,
    author = "Anacleto, M. A. and Campos, J. A. V. and Brito, F. A. and Passos, E.",
    title = "{Quasinormal modes and shadow of a Schwarzschild black hole with GUP}",
    eprint = "2108.04998",
    archivePrefix = "arXiv",
    primaryClass = "gr-qc",
    doi = "10.1016/j.aop.2021.168662",
    journal = "Annals Phys.",
    volume = "434",
    pages = "168662",
    year = "2021"
}

@article{Blanchette:2021vid,
    author = "Blanchette, Keagan and Das, Saurya and Rastgoo, Saeed",
    title = "{Effective GUP-modified Raychaudhuri equation and black hole singularity: four models}",
    eprint = "2105.11511",
    archivePrefix = "arXiv",
    primaryClass = "gr-qc",
    doi = "10.1007/JHEP09(2021)062",
    journal = "JHEP",
    volume = "09",
    pages = "062",
    year = "2021"
}

@article{Bosso:2023aht,
    author = "Bosso, Pasquale and Luciano, Giuseppe Gaetano and Petruzziello, Luciano and Wagner, Fabian",
    title = "{30 years in: Quo vadis generalized uncertainty principle?}",
    eprint = "2305.16193",
    archivePrefix = "arXiv",
    primaryClass = "gr-qc",
    doi = "10.1088/1361-6382/acf021",
    journal = "Class. Quant. Grav.",
    volume = "40",
    number = "19",
    pages = "195014",
    year = "2023"
}

@inproceedings{Ali:2010yn,
    author = "Ali, Ahmed Farag and Das, Saurya and Vagenas, Elias C.",
    title = "{The Generalized Uncertainty Principle and Quantum Gravity Phenomenology}",
    booktitle = "{12th Marcel Grossmann Meeting on General Relativity}",
    eprint = "1001.2642",
    archivePrefix = "arXiv",
    primaryClass = "hep-th",
    doi = "10.1142/9789814374552_0492",
    pages = "2407--2409",
    month = "1",
    year = "2010"
}

@article{GRAVITY:2021xju,
    author = "Abuter, R. and others",
    collaboration = "GRAVITY",
    title = "{Mass distribution in the Galactic Center based on interferometric astrometry of multiple stellar orbits}",
    eprint = "2112.07478",
    archivePrefix = "arXiv",
    primaryClass = "astro-ph.GA",
    doi = "10.1051/0004-6361/202142465",
    journal = "Astron. Astrophys.",
    volume = "657",
    pages = "L12",
    year = "2022"
}

@article{EventHorizonTelescope:2019dse,
    author = "Akiyama, Kazunori and others",
    collaboration = "Event Horizon Telescope",
    title = "{First M87 Event Horizon Telescope Results. I. The Shadow of the Supermassive Black Hole}",
    eprint = "1906.11238",
    archivePrefix = "arXiv",
    primaryClass = "astro-ph.GA",
    doi = "10.3847/2041-8213/ab0ec7",
    journal = "Astrophys. J. Lett.",
    volume = "875",
    pages = "L1",
    year = "2019"
}

@article{Islam:2022wck,
    author = "Islam, Shafqat Ul and Kumar, Jitendra and Kumar Walia, Rahul and Ghosh, Sushant G.",
    title = "{Investigating Loop Quantum Gravity with Event Horizon Telescope Observations of the Effects of Rotating Black Holes}",
    eprint = "2211.06653",
    archivePrefix = "arXiv",
    primaryClass = "gr-qc",
    doi = "10.3847/1538-4357/aca411",
    journal = "Astrophys. J.",
    volume = "943",
    number = "1",
    pages = "22",
    year = "2023"
}

@article{Devi:2021ctm,
    author = "Devi, Saraswati and S., Abhinove Nagarajan and Chakrabarti, Sayan and Majhi, Bibhas Ranjan",
    title = "{Shadow of quantum extended Kruskal black hole and its super-radiance property}",
    eprint = "2105.11847",
    archivePrefix = "arXiv",
    primaryClass = "gr-qc",
    doi = "10.1016/j.dark.2023.101173",
    journal = "Phys. Dark Univ.",
    volume = "39",
    pages = "101173",
    year = "2023"
}

@article{Corichi:2016nkp,
    author = "Corichi, Alejandro and Olmedo, Javier and Rastgoo, Saeed",
    archivePrefix = "arXiv",
    doi = "10.1103/PhysRevD.94.084050",
    eprint = "1608.06246",
    journal = "Phys.Rev.D",
    number = "8",
    pages = "084050",
    primaryClass = "gr-qc",
    title = "Callan-{G}iddings-{H}arvey-{S}trominger vacuum in loop quantum gravity and singularity resolution",
    volume = "94",
    year = "2016"
}

@article{Gambini:2009vp,
    author = "Gambini, Rodolfo and Pullin, Jorge and Rastgoo, Saeed",
    archivePrefix = "arXiv",
    doi = "10.1088/0264-9381/27/2/025002",
    eprint = "0909.0459",
    journal = "Class.Quant.Grav.",
    pages = "025002",
    primaryClass = "gr-qc",
    reportNumber = "LSU-REL-090209",
    title = "New variables for 1+1 dimensional gravity",
    volume = "27",
    year = "2010"
}

@article{Bosso:2020ztk,
    author = "Bosso, Pasquale and Obreg\'on, Octavio and Rastgoo, Saeed and Yupanqui, Wilfredo",
    title = "{Deformed algebra and the effective dynamics of the interior of black holes}",
    eprint = "2012.04795",
    archivePrefix = "arXiv",
    primaryClass = "gr-qc",
    doi = "10.1088/1361-6382/ac025f",
    journal = "Class. Quant. Grav.",
    volume = "38",
    number = "14",
    pages = "145006",
    year = "2021"
}

@article{Rastgoo:2022mks,
    author = "Rastgoo, Saeed and Das, Saurya",
    title = "{Probing the Interior of the Schwarzschild Black Hole Using Congruences: LQG vs. GUP}",
    eprint = "2205.03799",
    archivePrefix = "arXiv",
    primaryClass = "gr-qc",
    doi = "10.3390/universe8070349",
    journal = "Universe",
    volume = "8",
    number = "7",
    pages = "349",
    year = "2022"
}

@article{Bosso:2023fnb,
    author = "Bosso, Pasquale and Obreg\'on, Octavio and Rastgoo, Saeed and Yupanqui, Wilfredo",
    title = "{Black hole interior quantization: a minimal uncertainty approach}",
    eprint = "2310.04600",
    archivePrefix = "arXiv",
    primaryClass = "gr-qc",
    month = "10",
    year = "2023"
}

@article{Kempf:1994su,
    author = "Kempf, Achim and Mangano, Gianpiero and Mann, Robert B.",
    title = "{Hilbert space representation of the minimal length uncertainty relation}",
    eprint = "hep-th/9412167",
    archivePrefix = "arXiv",
    reportNumber = "DAMTP-94-105",
    doi = "10.1103/PhysRevD.52.1108",
    journal = "Phys. Rev. D",
    volume = "52",
    pages = "1108--1118",
    year = "1995"
}

@article{Perlick:2021aok,
    author = "Perlick, Volker and Tsupko, Oleg Yu.",
    title = "{Calculating black hole shadows: Review of analytical studies}",
    eprint = "2105.07101",
    archivePrefix = "arXiv",
    primaryClass = "gr-qc",
    doi = "10.1016/j.physrep.2021.10.004",
    journal = "Phys. Rept.",
    volume = "947",
    pages = "1--39",
    year = "2022"
}

@article{Fragomeno:2024tlh,
    author = "Fragomeno, Federica and Gingrich, Douglas M. and Hergott, Samantha and Rastgoo, Saeed and Vienneau, Evan",
    title = "{A generalized uncertainty-inspired quantum black hole}",
    eprint = "2406.03909",
    archivePrefix = "arXiv",
    primaryClass = "gr-qc",
    doi = "10.1103/PhysRevD.111.024048",
    journal = "Phys. Rev. D",
    volume = "111",
    number = "2",
    pages = "024048",
    year = "2025"
}

@article{Gingrich:2024mgk,
    author = "Gingrich, Douglas M. and Rastgoo, Saeed",
    title = "{Geometry of a generalized uncertainty-inspired spacetime}",
    eprint = "2412.08004",
    archivePrefix = "arXiv",
    primaryClass = "gr-qc",
    doi = "10.1103/PhysRevD.111.104017",
    journal = "Phys. Rev. D",
    volume = "111",
    number = "10",
    pages = "104017",
    year = "2025"
}

@article{Kumar:2019uwi,
    author = "Kumar, Utkarsh and Panda, Sukanta and Patel, Avani",
    title = "{Blackhole in nonlocal gravity: comparing metric from Newmann{\textendash}Janis algorithm with slowly rotating solution}",
    eprint = "1906.11714",
    archivePrefix = "arXiv",
    primaryClass = "gr-qc",
    doi = "10.1140/epjc/s10052-020-8182-5",
    journal = "Eur. Phys. J. C",
    volume = "80",
    number = "7",
    pages = "614",
    year = "2020"
}

@article{Balali:2024mtt,
    author = "Balali, A. El and Benali, M. and Oualaid, M.",
    title = "{Deflection angle and shadow of slowly rotating black holes in galactic nuclei}",
    eprint = "2401.02341",
    archivePrefix = "arXiv",
    primaryClass = "gr-qc",
    doi = "10.1007/s10714-024-03205-z",
    journal = "Gen. Rel. Grav.",
    volume = "56",
    number = "2",
    pages = "21",
    year = "2024"
}

@article{Nicolini:2005vd,
    author = "Nicolini, Piero and Smailagic, Anais and Spallucci, Euro",
    title = "{Noncommutative geometry inspired Schwarzschild black hole}",
    eprint = "gr-qc/0510112",
    archivePrefix = "arXiv",
    doi = "10.1016/j.physletb.2005.11.004",
    journal = "Phys. Lett. B",
    volume = "632",
    pages = "547--551",
    year = "2006"
}

@article{Ashtekar:2018cay,
    author = "Ashtekar, Abhay and Olmedo, Javier and Singh, Parampreet",
    title = "{Quantum extension of the Kruskal spacetime}",
    eprint = "1806.02406",
    archivePrefix = "arXiv",
    primaryClass = "gr-qc",
    doi = "10.1103/PhysRevD.98.126003",
    journal = "Phys. Rev. D",
    volume = "98",
    number = "12",
    pages = "126003",
    year = "2018"
}

@article{Newman:1965tw,
    author = "Newman, E. T. and Janis, A. I.",
    title = "{Note on the Kerr spinning particle metric}",
    doi = "10.1063/1.1704350",
    journal = "J. Math. Phys.",
    volume = "6",
    pages = "915--917",
    year = "1965"
}

@article{Newman:1961qr,
    author = "Newman, Ezra and Penrose, Roger",
    title = "{An Approach to gravitational radiation by a method of spin coefficients}",
    doi = "10.1063/1.1724257",
    journal = "J. Math. Phys.",
    volume = "3",
    pages = "566--578",
    year = "1962"
}

@article{HansenYunes2013,
    author = "Hansen, Devin and Yunes, Nicolas",
    title = "{Applicability of the Newman-Janis algorithm to black hole solutions of modified gravity theories}",
    eprint = "1308.6631",
    archivePrefix = "arXiv",
    primaryClass = "gr-qc",
    doi = "10.1103/PhysRevD.88.104020",
    journal = "Phys. Rev. D",
    volume = "88",
    pages = "104020",
    year = "2013"
}

@article{KamenshchikPetriakova2023,
    author = "Kamenshchik, Alexander and Petriakova, Polina",
    title = "{Regular rotating black hole: To Kerr or not to Kerr?}",
    eprint = "2211.04542",
    archivePrefix = "arXiv",
    primaryClass = "gr-qc",
    doi = "10.1103/PhysRevD.107.124020",
    journal = "Phys. Rev. D",
    volume = "107",
    pages = "124020",
    year = "2023"
}

@article{ShaoChenChen2021,
    author = "Shao, Wei-Hsiang and Chen, Che-Yu and Chen, Pisin",
    title = "{Generating rotating spacetime in Ricci-based gravity: naked singularity as a black hole mimicker}",
    eprint = "2011.07763",
    archivePrefix = "arXiv",
    primaryClass = "gr-qc",
    doi = "10.1088/1475-7516/2021/03/041",
    journal = "JCAP",
    volume = "03",
    pages = "041",
    year = "2021"
}

@article{NevesSaa2014,
    author = "Neves, Joao C. S. and Saa, Alberto",
    title = "{Regular rotating black holes and the weak energy condition}",
    eprint = "1404.6310",
    archivePrefix = "arXiv",
    primaryClass = "gr-qc",
    doi = "10.1016/j.physletb.2014.05.026",
    journal = "Phys. Lett. B",
    volume = "734",
    pages = "44--48",
    year = "2014"
}

@book{bambi2023regular,
  editor    = {Cosimo Bambi},
  title     = {Regular Black Holes: Towards a New Paradigm of Gravitational Collapse},
  year      = {2023},
  publisher = {Springer},
  isbn      = {978-981-99159-5-8}
}

@article{Hergott:2022hjm,
    author = "Hergott, Samantha and Husain, Viqar and Rastgoo, Saeed",
    title = "{Model metrics for quantum black hole evolution: Gravitational collapse, singularity resolution, and transient horizons}",
    eprint = "2206.06425",
    archivePrefix = "arXiv",
    primaryClass = "gr-qc",
    doi = "10.1103/PhysRevD.106.046012",
    journal = "Phys. Rev. D",
    volume = "106",
    number = "4",
    pages = "046012",
    year = "2022"
}

@article{Hergott:2025elg,
    author = "Hergott, Samantha and Husain, Viqar and Rastgoo, Saeed",
    title = "{Dynamical model for black hole to white hole transitions}",
    eprint = "2505.15096",
    archivePrefix = "arXiv",
    primaryClass = "gr-qc",
    doi = "10.1103/8m7g-zz3k",
    journal = "Phys. Rev. D",
    volume = "113",
    number = "2",
    pages = "024049",
    year = "2026"
}

@article{Amati:1988tn,
    author = "Amati, D. and Ciafaloni, M. and Veneziano, G.",
    title = "{Can Space-Time Be Probed Below the String Size?}",
    journal = "Phys. Lett. B",
    volume = "216",
    pages = "41--47",
    year = "1989"
}

@article{Gross:1987ar,
    author = "Gross, David J. and Mende, Paul F.",
    title = "{String Theory Beyond the Planck Scale}",
    journal = "Nucl. Phys. B",
    volume = "303",
    pages = "407--454",
    year = "1988"
}

@article{Snyder:1946qz,
    author = "Snyder, Hartland S.",
    title = "{Quantized space-time}",
    journal = "Phys. Rev.",
    volume = "71",
    pages = "38--41",
    year = "1947"
}

@book{Connes:1994yd,
    author = "Connes, Alain",
    title = "{Noncommutative geometry}",
    publisher = "Academic Press",
    address = "San Diego, CA",
    year = "1994"
}

@article{Rovelli:1994ge,
    author = "Rovelli, Carlo and Smolin, Lee",
    title = "{Discreteness of area and volume in quantum gravity}",
    journal = "Nucl. Phys. B",
    volume = "442",
    pages = "593--622",
    year = "1995",
    eprint = "gr-qc/9411005"
}

@article{Maggiore:1993rv,
    author = "Maggiore, Michele",
    title = "{A Generalized uncertainty principle in quantum gravity}",
    journal = "Phys. Lett. B",
    volume = "304",
    pages = "65--69",
    year = "1993",
    eprint = "hep-th/9301067"
}

@article{Bonanno:2000ep,
    author = "Bonanno, Alfio and Reuter, Martin",
    title = "{Renormalization group improved black hole space-times}",
    journal = "Phys. Rev. D",
    volume = "62",
    pages = "043008",
    year = "2000",
    doi = "10.1103/PhysRevD.62.043008",
    eprint = "hep-th/0002196"
}

@article{Reuter:2010xbd,
    author = "Reuter, Martin and Tuiran, E.",
    title = "{Quantum Gravity Effects in Rotating Black Hole Spacetimes}",
    journal = "Phys. Rev. D",
    volume = "83",
    pages = "044041",
    year = "2011",
    doi = "10.1103/PhysRevD.83.044041",
    eprint = "1009.3528"
}

@article{Wald:1993nt,
    author = "Wald, Robert M.",
    title = "{Black hole entropy is the Noether charge}",
    journal = "Phys. Rev. D",
    volume = "48",
    pages = "3427--3431",
    year = "1993",
    doi = "10.1103/PhysRevD.48.R3427",
    eprint = "gr-qc/9307038"
}

@article{Kaul:2000iq,
    author = "Kaul, Romesh K. and Majumdar, Parthasarathi",
    title = "{Logarithmic correction to the Bekenstein-Hawking entropy}",
    journal = "Phys. Rev. Lett.",
    volume = "84",
    pages = "5255--5257",
    year = "2000",
    doi = "10.1103/PhysRevLett.84.5255",
    eprint = "gr-qc/0002040"
}

@article{Carlip:2000nv,
    author = "Carlip, Steven",
    title = "{Logarithmic corrections to black hole entropy from the Cardy formula}",
    journal = "Class. Quant. Grav.",
    volume = "17",
    pages = "4175--4186",
    year = "2000",
    doi = "10.1088/0264-9381/17/20/302",
    eprint = "gr-qc/0005017"
}

@article{Adler:2001vs,
    author = "Adler, Ronald J. and Chen, Pisin and Santiago, David I.",
    title = "{The Generalized uncertainty principle and black hole remnants}",
    journal = "Gen. Rel. Grav.",
    volume = "33",
    pages = "2101--2108",
    year = "2001",
    doi = "10.1023/A:1015281430411",
    eprint = "gr-qc/0106080"
}

\end{document}